\documentclass[useAMS,usenatbib]{mn2e}
\usepackage{rotating}     
\usepackage{graphicx}
\usepackage{amsmath}
\usepackage{amsfonts}
\usepackage{amssymb}
\usepackage{natbib}
\usepackage{color}

\newcommand{\mbi}[1]{\textbf{\emph{#1}}}

\usepackage{xspace}

\def\kpch{\mbox{$h^{-1}$\,kpc}}

\def\Mpc{\mbox{Mpc}}

\def\Mpch{\mbox{$h^{-1}$\,Mpc}}

\def\M200{\mbox{$M_{\rm 200}$}}

\def\Msunh{\mbox{$h^{-1}\,{\rm M}_\odot$}}

\def\R200{\mbox{$R_{\rm 200}$}}

\def\V200{\mbox{$V_{\rm 200}$}}

\title[The cosmic web of the Local Universe]{The cosmic web of the Local Universe: cosmic variance, 
matter content and its relation to galaxy morphology}

\author[Nuza et al.]{Sebasti\'an\,E.\,Nuza\thanks{E-mail: snuza@aip.de}, 
Francisco-Shu Kitaura\thanks{E-mail: fkitaura@aip.de, Karl-Schwarzschild-fellow}, 
Steffen He\ss\thanks{E-mail: shess@aip.de}, Noam I. Libeskind 
\newauthor
and Volker M\"uller\\
Leibniz-Institut f\"ur Astrophysik Potsdam, An der Sternwarte 16, 14482 Potsdam, Germany\\
}

\date{}

\begin{document}

\maketitle

\begin{abstract}
We present, for the first time, a Local Universe (LU) characterization using high precision 
constrained $N$-body simulations based on self-consistent phase-space reconstructions of the 
large-scale structure in the Two-Micron All-Sky Galaxy Redshift Survey. We analyse whether 
we live in a special cosmic web environment by estimating cosmic variance from a set of 
unconstrained $\Lambda$CDM simulations as a function of distance to random observers. By computing 
volume and mass filling fractions for voids, sheets, filaments and knots, we find that the LU 
displays a typical scatter of about $1\sigma$ at scales $r\gtrsim15\,\Mpch$, in agreement with 
$\Lambda$CDM, converging  to a fair unbiased sample when considering spheres of about $60\,\Mpch$ radius. 
Additionally, we compute the matter density profile of the LU and found a reasonable agreement with 
the estimates of \cite{Karachentsev12} only when considering the contribution of dark haloes. 
This indicates that observational estimates may be biased towards low density values. 
As a first application of our reconstruction, we investigate the likelihood of 
different galaxy morphological types to inhabit certain cosmic web environments. 
In particular, we find that, irrespective of the method used to define the web, either 
based on the density or the peculiar velocity field, elliptical galaxies show a clear tendency 
to preferentially reside in clusters as opposed to voids 
(up to a level of $5.3\sigma$ and $9.8\sigma$ respectively) and conversely for spiral galaxies 
(up to a level of $5.6\sigma$ and $5.4\sigma$ respectively). 
These findings are compatible with previous works, albeit at higher confidence levels.
\end{abstract}  


\section{Introduction}
\label{sec:intro}

\begin{figure*}
\begin{center}
  \includegraphics[width=0.98\textwidth]{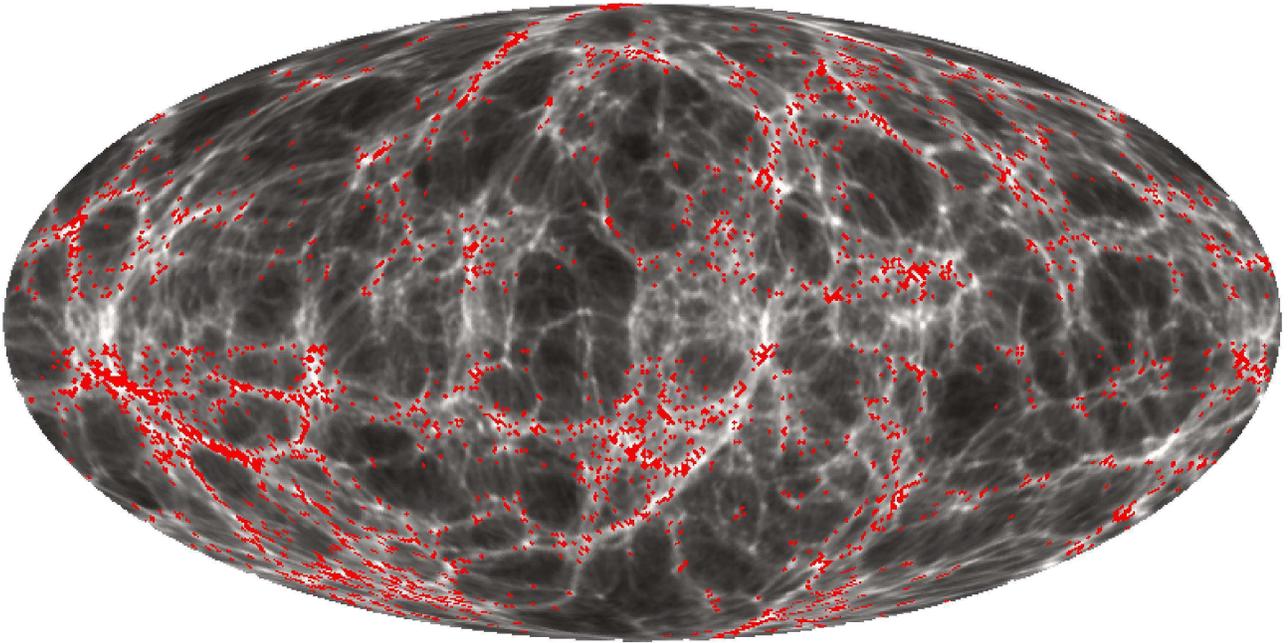}
  \end{center}
  \caption{Mollview projection in Galactic coordinates of all galaxies in the 2MRS catalogue (red dots) 
  at distances of $50-60\,\Mpch$ generated using HEALPix \citep{Gorski05}. The grey colour scale shows 
  the DM density field of our best-correlated $N$-body reconstruction. The initial conditions 
  of the reconstruction have been produced using the \textsc{kigen}-code (see Section~\ref{sec:csims}). 
  Note that the empty region around the Galactic plane corresponds to the {\it Zone of Avoidance}.
  }
  \label{fig:sky_proj}
\end{figure*}

The location of our place in the Universe was found to be quite particular in a number 
of studies. Our local environment within a distance of about $50\,\Mpc$ 
appears to be less dense than the expected mean density of the Universe 
\citep[e.g.,][]{Vennik84,Tully87,Magtesian88,Bahcall00,Crook07,Makarov11,Karachentsev12}. 
Additionally, a prominent low density region, the so-called Local Void, is found to be 
located in our immediate vicinity \citep{TullyFisher87}. At larger distances, 
it has been found that most of the nearby clusters lie on an approximately ring-like 
structure that surrounds our location. This feature led \cite{deVaucouleurs53} to 
propose the super-galactic coordinate system in the plane of this structure 
to enhance this curious pattern \citep[][]{deVaucouleurs76,deVaucouleurs91}. 

After considering these peculiarities a natural question 
arises: is our location in the Universe special? 
More specifically: how likely are the prevailing structures in our Local Universe (LU)? 
Are these local patterns capable of challenging the $\Lambda$CDM cosmology?
To answer these questions we first need to define those cosmic patterns in a quantitative way.
The distribution of galaxies as obtained by galaxy redshift surveys, such as the Two-Micron 
All-Sky Galaxy Redshift Survey \citep[2MRS;][]{Huchra12} which we consider in this work, reveals the 
existence of an intricate network of interconnected structures comprising a web of walls, 
filaments, galaxy clusters and voids \citep[][]{Zeldovich70,Peebles80,Bond96}. 
According to the picture introduced by \citet{Zeldovich70} such a cosmic web naturally 
arises as a result of gravitational collapse. Early numerical simulations
(e.g., \citealt{Doroshkevich80,Davis85}) have confirmed this scenario. 
These different environments are characterized by their distinctive dynamical nature: voids 
are expanding low-density regions whereas clusters are collapsing dense structures residing 
at the intersection of elongated filaments. Similarly, filaments are `chains' of galaxies 
being constantly stretched across their major axes that are located where two-dimensional 
expanding sheets meet.

Ever since the pioneering works of \citet{Fry78} and \citet{White78} the hierarchical 
structure formation paradigm has been established in cosmology. According to this picture, 
structures are formed in a hierarchical process, in which smaller objects merge to form 
larger ones. Such a process will predominantly occur in higher density environments, where 
mergers are more likely to happen. Numerical simulations 
have long since shown that mergers or tidal interactions can destroy galactic discs converting 
spiral or irregular galaxies into elliptical and S0 galaxies \citep{Toomre72,Farouki81}. 
Therefore, this mechanism can naturally generate a morphological segregation of galaxies 
as the density of the environment increases. 
In fact, such a morphology-density relation was already discovered by \citet{Oemler74} and 
\citet{Dressler80}, showing that star-forming, disc-dominated galaxies 
preferentially reside in lower density regimes as opposed to elliptical ones. 

A large number of works have further investigated the relation
between the environment and galaxy properties in the LU, such as 
morphological type, stellar mass,
(specific) star formation rate, colour and luminosity
(e.g., \citealt{Balogh01,Balogh04,Hogg04,Kauffmann04,Tanaka04,Blanton05,
Croton05,Alonso06,Baldry06,Martinez06,Weinmann06,Einasto07,Park07,Ball08,
vanderWel08,Deng11,Tempel11,Zandivarez11,Alonso12,Wetzel12,Lackner13}).
Most of these studies have focused on the high density
regime, which can relatively easily be determined through the
local number density of galaxies. 
Additionally, other authors have focused in the classification of
filamentary-like structures
\citep[e.g.,][]{Sousbie08,Stoica10,Sousbie11,Smith12,Beygu13,Tempel14b} and voids 
\citep[e.g.,][]{Mueller00,Hoyle04,Ceccarelli06,Kreckel12,Lietzen12,Pan12,Sutter12,
Ceccarelli13,Sutter13,Tavasoli13} as a way to shed light on the galaxy 
formation process in these environments.

Despite these efforts, a comprehensive study of the local environment with respect 
to the nonlinear cosmic web is still missing. 
In this respect, some remarkable attempts have been presented in the 
literature. For instance, \cite{LeeLee08} used a LU reconstruction 
based on a linear Wiener filter method that, however, tends to smooth out structures 
at scales of the order of $10\,\Mpch$ (see \citealt{Erdogdu04,Erdogdu06}, as well as the 
discussion in \citealt{Kitaura09}). \citet{Nuza10} studied the properties 
of simulated galaxy populations in a simpler reconstruction of the LU, with the additional 
complication of including 
gas physics. Another interesting example is the work of \citet{MunozCuartas11} who performed a 
halo-based reconstruction of the local web by convolving the information of 
unconstrained $N$-body simulations with a group galaxy catalog. Alternatively, 
\cite{AragonCalvo12} performed an ensemble of randomly seeded $N$-body simulations 
displaying the same large-scale structure, but different small-scale perturbations, 
aiming at studying the statistics of haloes as a function of environment.

In this context, it is mandatory to revise the environmental studies with more accurate 
reconstructions of the large-scale structure. We therefore extend such works using 
the recently performed high precision constrained 
simulations that correlate with 2MRS galaxies down to a few Mpc scales \citep{Hess13}. 
These simulations are based 
on the first self-consistent phase-space reconstruction method of the primordial fluctuations 
corresponding to a set of matter tracers (the \textsc{kigen}-code: \citealt{Kitaura13}). At the same 
time, the unprecedented accuracy of our simulations, permit us to characterize the dark matter (DM) content 
within the LU and compare with observational estimates. 

In this work, we present a systematic study of the cosmic web, as predicted both by random simulations 
and precise reconstructions of the LU after studying the eigenvalues of the tidal field tensor 
of the density field \citep{Hahn07a,Forero09}. As will be explained further, while classifying 
the web, we will use the information contained in the nonlinear and 
linear reconstructed density fields to assess the robustness of the measurements. It is worth 
noting, however, that the linear overdensity will be estimated only by means of the nonlinear 
velocities of the reconstruction which are known to be more linear than the density field 
(see e.g., \citealt{Zaroubi99}; \citealt{Kitaura12a} and references therein). Moreover, the reconstructed 
velocity field yields a complementary environmental view that is based on the kinematics of the LU.

In summary, the aim of the present work is twofold. In the first place, we aim at studying 
the LU matter content as well as the impact of cosmic variance as a function of distance to the observer. 
In this way, we will be able to assess the scale at which our LU becomes a {\it fair sample}. 
To do so, we characterize the cosmic web of the LU by using high precision constrained $N$-body 
simulations with two different approaches. In particular, we compute the volume and mass filling 
fractions (VFFs and MFFs, respectively) of different cosmic web environments. 
Then, we compare these statistics with those corresponding to a set of unconstrained randomly 
seeded $\Lambda$CDM simulations. 
Secondly, we want to benefit from the high level of accuracy of our reconstructions 
to measure the correlation between galaxy morphology and their location within 
the cosmic web, presenting a first application of the LU density field 
estimation for the galaxy population inhabiting the nearby Universe.

\begin{figure*}
  \centering	
  \includegraphics[width=0.5\textwidth]{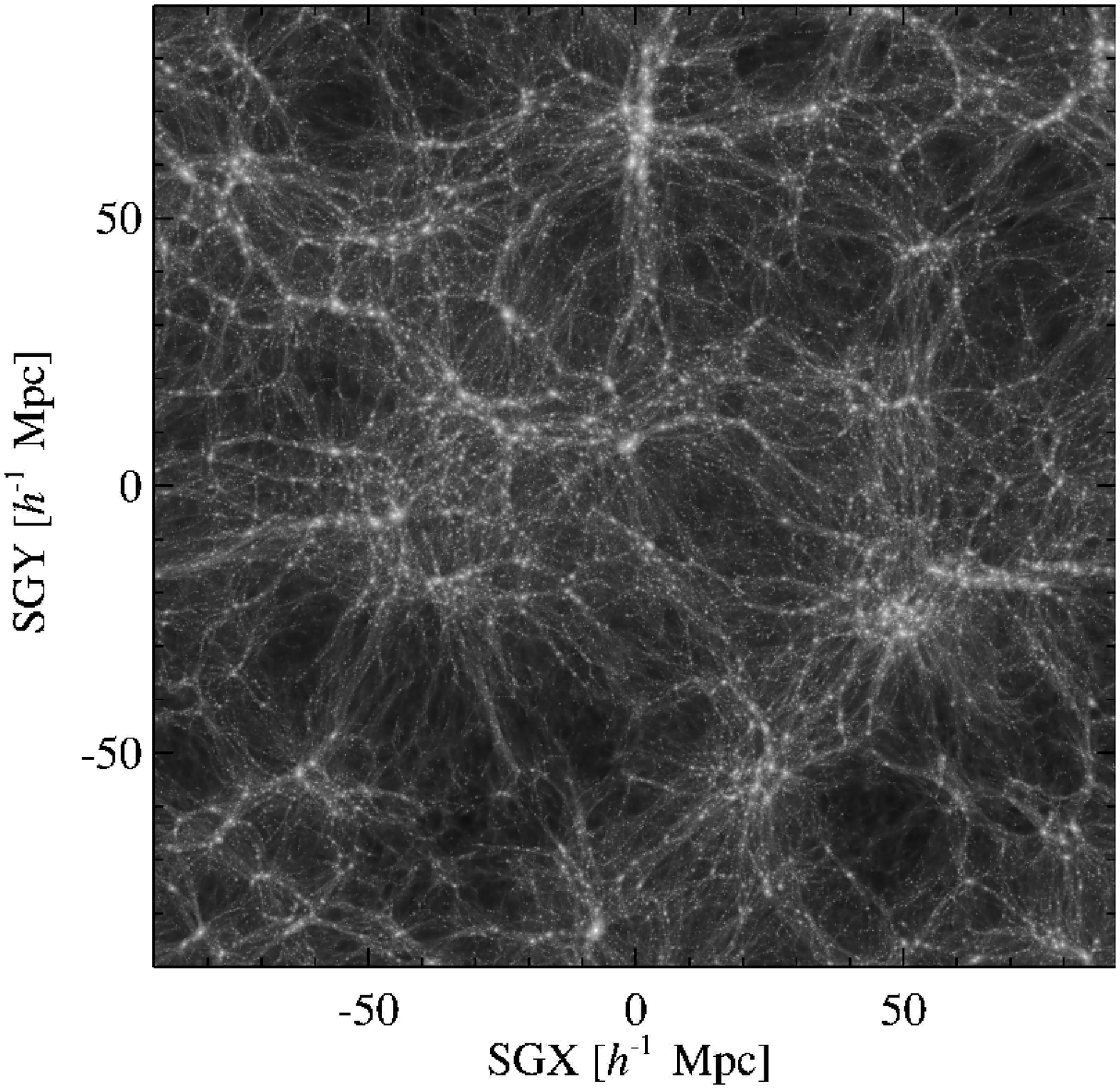}\includegraphics[width=0.506\textwidth]{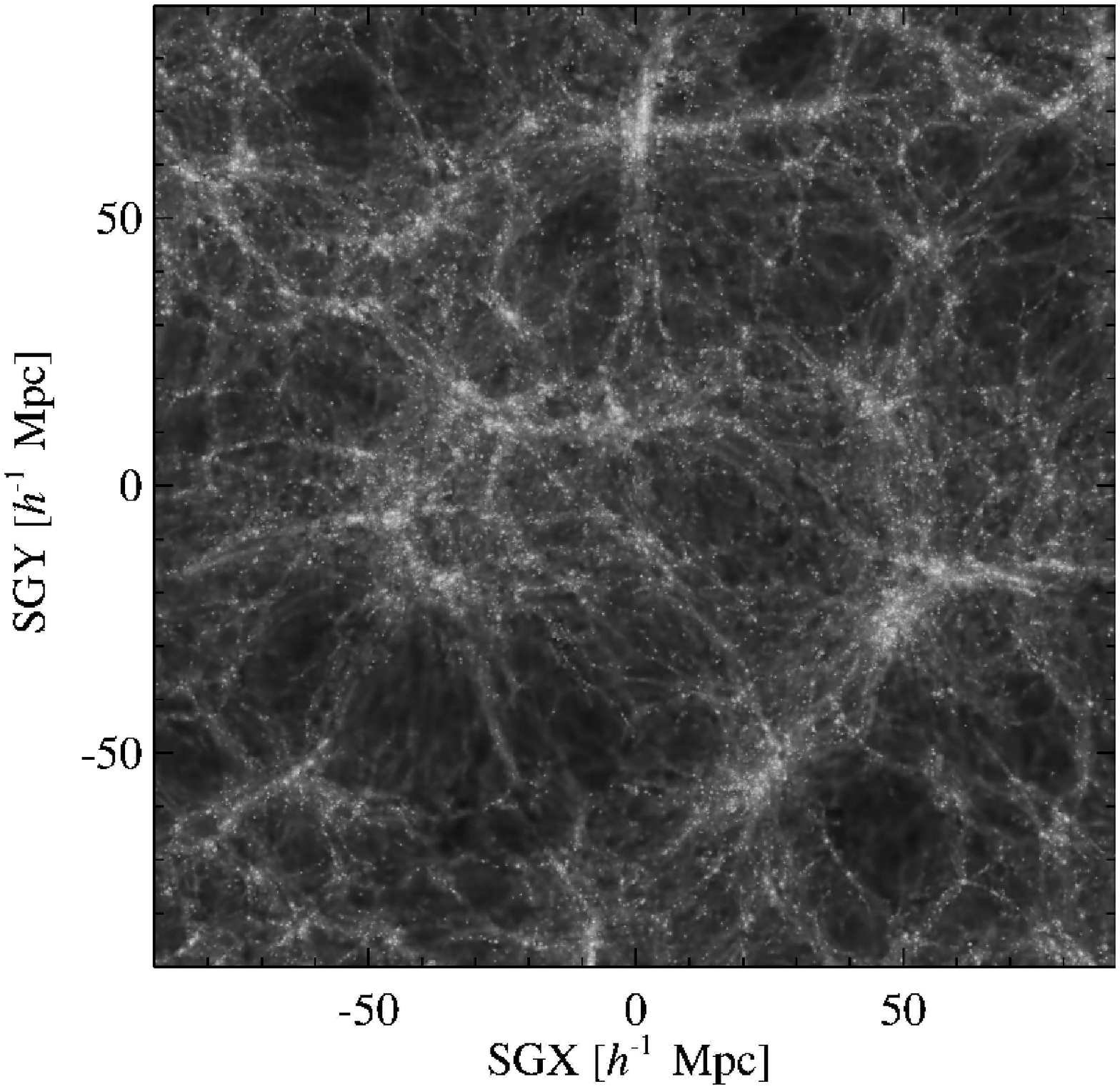}
 \caption{DM overdensity map in a slice of $22.5\,\Mpch$ centred on the position 
 of the observer for our best LU reconstruction in real and redshift space (left and right panels respectively; see \citealt{Hess13}).
 Units are in supergalactic coordinates.}
  \label{fig:DM_overdens} 
\end{figure*}

This paper is structured as follows. First, in
Section~\ref{sec:input}, we introduce the dataset used to generate the ICs, 
then we briefly present the method applied to reconstruct the large-scale structure, 
as well as some details of our constrained simulations. In Section~\ref{sec:DM_content} 
we compute the matter density parameter of the LU as a function of distance 
to the observer and compare with recent observational estimates. 
In Section~\ref{sec:cosweb} we discuss the large-scale structure 
classification methods used to define the local web and the 
resulting statistics in each case. 
In Section~\ref{sec:cosmic_variance} 
we assess the cosmic variance level of the LU in comparison with the 
expectations of $\Lambda$CDM. In Section~\ref{sec:env} we 
study the galaxy morphology/cosmic web correlation, for our web 
classifications, as a first application 
of the LU reconstruction. Finally, in Section~\ref{sec:conc}, 
we present a summary and give our conclusions.

\section{Input data}
\label{sec:input}

\subsection{The 2MRS survey}
\label{sec:2mrs}

Our study is based on the Two-Micron All-Sky Galaxy Redshift Survey
(2MRS) K$_{\rm s}=11.75$ catalogue presented by \cite{Huchra12}.  The
unprecedented sky coverage ($91\%$) and uniform completeness ($97\%$)
in the 2MRS galaxy catalogue are ideal to probe the characteristics of
local structures.  Observations are only limited by the {\it Zone of
  Avoidance}.  We note that we could treat the mask in a
self-consistent way within the reconstruction process \citep[see
  e.g.,][]{Jasche2010}. However, the small number of galaxies
affected by the mask, permits us to correct for this effect using
the data augmented catalog by \cite{Erdogdu06}. This is done adding random galaxies
drawn from the corresponding adjacent strips of the survey
\citep{Yahil91}. We consider the volume within a box of
$180\,\Mpch$ on a side with its centre located at the observer's
position. As a result, our final sample comprises 31,017 galaxies, which 
corresponds to about 76\% of the 2MRS survey. 
In this way, we are able to avoid the steep decrease of the
radial selection function at the edge of the survey containing less
than one quarter of the available galaxies. This also permits us to
minimize the Kaiser-rocket effect \citep{Branchini12}.

\subsection{The reconstruction method: a self-consistent phase-space forward approach}
\label{sec:csims}

To obtain the full nonlinear phase-space distribution of a set of matter tracers in redshift space, 
we rely on the \textsc{kigen}-code \citep[][]{Kitaura13,Kitaura12}, which is the first self-consistent, 
phase-space, forward reconstruction approach. Other remarkable forward approaches have been 
developed \citep[see][]{Jasche13,Wang13}, although without including a self-consistent estimation 
of peculiar velocities, as we do here. 

The advantage of the \textsc{kigen}-code is manifold, as it includes:
\begin{enumerate}
\item An accurate gravitational collapse model on Mpc scales combining  second order LPT (2LPT) at 
large scales with the spherical collapse model at cluster scales 
\citep[dubbed `Augmented' Lagrangian perturbation theory, hereafter ALPT;][]{KitauraHess13} 
that suppress shell-crossing in the high-density regime and improves the description of 
filaments at lower densities \citep[see also][]{Neyrinck13}.\\

\item Nonlinear coherent and virialised redshift-space distortions built-in in the second step 
(likelihood comparison) of the \textsc{kigen}-code \citep[see][]{Hess13}, i.e. without artificially 
compressing the Fingers-of-God.\\

\item A nonlinear scale dependent bias in Lagrangian space to ensure that the reconstructed primordial 
density fields yield unbiased power spectra, as compared to the linear theoretical 
model (this happens in the first step of \textsc{kigen}). In particular, we use the exponential 
relation proposed by \citet[][]{Cen93} that is able to model the nonlinear scale-dependent 
bias as it was shown in \citet[][]{delatorre13}. We note that the joint treatment of peculiar 
velocities (see (i)) and biasing in the initial conditions is essential to break the degeneracy 
present in the two-point statistics.
\end{enumerate}

\subsection{Constrained $N$-body simulations}

In this work, we rely on constrained $N$-body simulations based on reconstructions performed 
with the \textsc{kigen}-code applied to the 2MRS survey \citep{Hess13}. 
In particular, we consider in our study a subset of 25 constrained simulations displaying the 
largest cross-correlations with the galaxy data. 
To estimate the degree of correlation between the reconstructed DM density field 
($\delta_{\rm DM}$) and galaxy overdensities ($\delta_{\rm G}$) we define the cross power-spectrum as
\begin{equation}	
XP(k)[\delta_{\rm DM},\delta_{\rm G}]\equiv
\frac{\langle|\hat{\delta}_{\rm DM}(\mbi k)\overline{\hat{\delta}_{\rm G}(\mbi k)}|\rangle}
{\sqrt{P_{\rm DM}(k)}\sqrt{P_{\rm G}(k)}}\,,
\end{equation}
\noindent where the ensemble brackets denote angular averaging and $P_{\rm DM}(k)$ 
and $P_{\rm G}(k)$ are the associated power spectra.
A cell-to-cell comparison of the logarithmic density fields in configuration space shows a 
typical Pearson coefficient of 74\% for a cell width of $1.4\,\Mpch$. 
The high degree of correlation between the 2MRS galaxy distribution and the reconstructed density field 
can be seen in Fig.~\ref{fig:sky_proj}. This plot shows a Mollview sky projection of all matter 
(grey scale) and galaxies (red dots) in our best-correlated real-space reconstruction 
at a distance of $50-60\,\Mpch$. A remarkable spatial coincidence between the galaxy tracers 
and the underlying density field can be observed.

The difference between real and redshift space for our LU reconstruction can be seen 
in Fig.~\ref{fig:DM_overdens} (left and right panels respectively). This figure demonstrates 
the squashing effect produced along the line-of-sight as a result of the coherent peculiar motions 
of galaxies. In this sense, it is important to perform the cosmic web analysis both in configuration and redshift 
space in order to check for differences in the results. 
These runs are complemented with an analogous set of DM-only simulations 
carried out using unconstrained random phases for the ICs adopting the same cosmology 
and simulation parameters. This set of unconstrained simulations will serve to assess 
the level of cosmic variance of the LU reconstructions. The reader is referred to 
\citet{Hess13} for further details.  

A flat $\Lambda$CDM model consistent with WMAP7 results \citep[][]{Komatsu11} was considered 
in these simulations, i.e. with  a matter density $\Omega_{\rm M}=0.272$, 
a cosmological constant density of $\Omega_{\Lambda}=0.728$, 
a baryon density of $\Omega_{\rm b}=0.046$,  a dimensionless Hubble constant $h=0.704$, 
an amplitude of mass fluctuations $\sigma_8=0.807$ and a scalar spectral index $n_{\rm s}=0.967$. 
The simulations were initialised at redshift $z=100$ and were evolved 
with the {\sc gadget-3} code \citep{Springel05,Springel08} using $384^3$ particles, 
which translates into a DM particle mass of $7.8\times10^9\,\Msunh$, 
and a comoving gravitational softening of $15\,\kpch$. Throughout this paper, 
we will use the density fields of the $N$-body simulations obtained from 
their corresponding DM distribution using a standard cloud-in-cell technique. 
The volume has been gridded with $128^3$ cells thus giving a cell side length 
of about $1.4\,\Mpch$.

DM haloes in the simulations are identified using the {\sc ahf} code \citep{KnollmannKnebe09} 
as spherical overdensities with a density $200$ times above the 
critical density of the universe. In this way, a halo sample with masses 
in the range $\sim$$10^{11}-10^{14}\,\Msunh$ was selected for our LU reconstructions 
\citep[see Fig. 10 of][for the mass function of our constrained realisations]{Hess13}. 
To test for resolution and cosmology effects, when haloes are considered, we also used a higher resolution 
reconstruction with $768^3$ particles within the Planck 
cosmology \citep[with $\Omega_{\rm M}=0.307$;][]{PLANCK13}, which results in a DM particle mass 
of $1.1\times10^9\,\Msunh$.

\begin{figure}
 \hspace{-0.5cm}
 \includegraphics[width=0.53\textwidth]{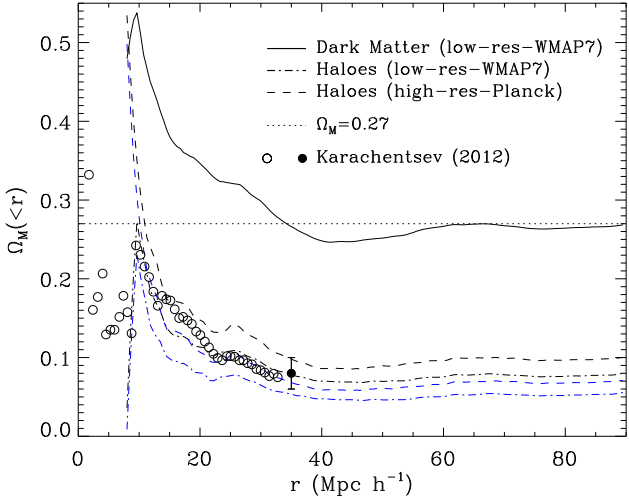}
 \caption{Matter density parameter as a function of radius for increasingly larger spheres as obtained 
 from our reconstruction of the LU. Shown are the results corresponding to the DM 
 as well as the halo contribution for virial halo mass 
 cuts of $M_{\rm vir}>10^{12}$ and $5\times10^{12}\,\Msunh$ (black and blue lines respectively).  
 Also shown are the observational results of \citet{Karachentsev12} which have been estimated from 
 an updated local galaxy sample (circles). The solid circle with error bars stands for the integrated 
 observational result within $50\,\Mpc$ ($\equiv35\,\Mpch$).} 
\label{fig:OmegaM_vs_r}
\end{figure}

\section{Matter content in the LU: haloes vs. field}
\label{sec:DM_content}

Throughout the years there have been claims that the LU may not be a fair
sample of the Universe after studying the contribution of galaxy groups to its
DM content. Typically these estimates place the local matter density
value in the range $\Omega_{\rm M,LU}\approx0.05-0.2$, i.e. below the
cosmological mean density by a factor of a few (see
e.g. \citealt{Karachentsev12} and references therein). 
These results have been sometimes interpreted as the consequence of an
underlying `missing DM problem' that could potentially be in conflict
with the $\Lambda$CDM cosmological model. In particular, the recent study of
\cite{Karachentsev12} presents an updated analysis of a sample of local
galaxies -- that includes dwarfs, pairs, triplets and larger groups --
favouring an estimate of $\Omega_{\rm M,LU} = 0.08 \pm 0.02$ for the volume
contained within a distance of $50\,$Mpc.

The amount of DM as a function of volume in our reconstructions can
be straightforwardly checked for consistency with observational
results. This is shown in Fig.~\ref{fig:OmegaM_vs_r} where the cumulative
matter density parameter is calculated as a function of distance to the
observer for our best (real-space) $N$-body reconstruction. It can be seen
that the DM density shows some modulation owing to the particular LU
realisation to finally converge to the mean cosmological value at a distance
of about $60\,\Mpch$, i.e. approximately $85\,$Mpc. The Local Void manifests
itself as a slight trough in the cumulative matter density for distances of
about $30-60\,\Mpch$. This clearly indicates that the observed DM
density estimates in the LU cannot be the result of cosmic variance as the
corresponding minimum value in our reconstruction is well 
above $\Omega_{\rm M}=0.1$.

The situation changes if we consider the contribution of dark haloes
alone.  The dot-dashed (dashed) lines show the density parameter
obtained from matter located within the virial radius of haloes for
virial mass cuts of $M_{\rm vir}>10^{12}$ and $5\times10^{12}\,\Msunh$
(black and blue lines respectively) for the low (high) resolution
$N$-body reconstruction. These cuts are roughly consistent with the
minimum virial mass value of galaxy groups with $N>3$ in the
\cite{Karachentsev12} sample as can be inferred from their
Fig. 2. According to this work, these systems are the main
contributors to the cosmic matter budget in the LU and have a median
virial mass of about $1.69\times10^{12}\,\Msunh$ \citep{Makarov11}. In
general, for the reconstructed halo samples, the predictions of both
constrained simulations using different matter density parameters 
($\Omega_{\rm M}=0.272$ and $\Omega_{\rm M}=0.307$, respectively) are
compatible with the observational estimates of \cite{Karachentsev12}
within the error bars.

\begin{figure*}
  \includegraphics[width=0.49\textwidth]{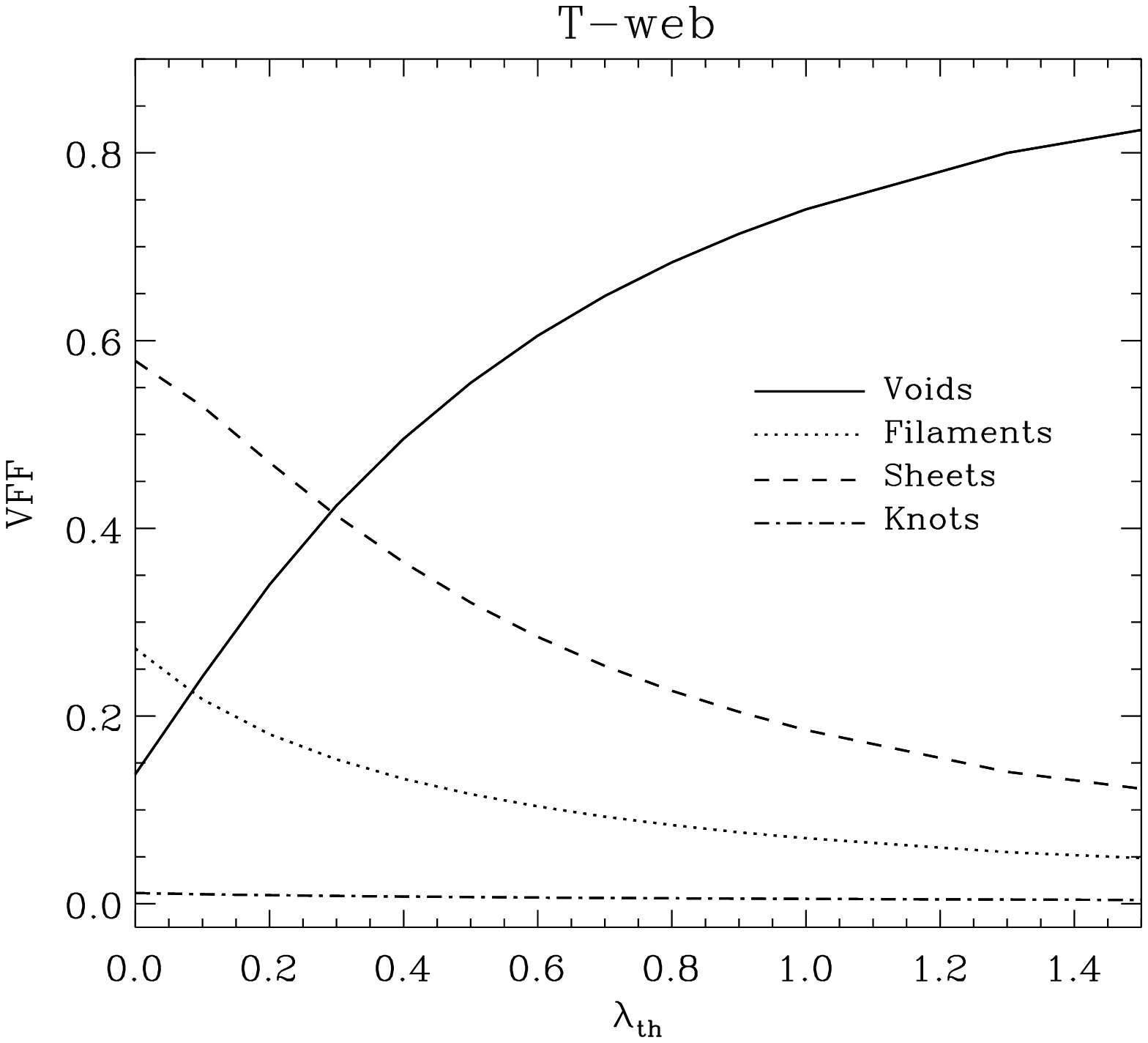}\includegraphics[width=0.49\textwidth]{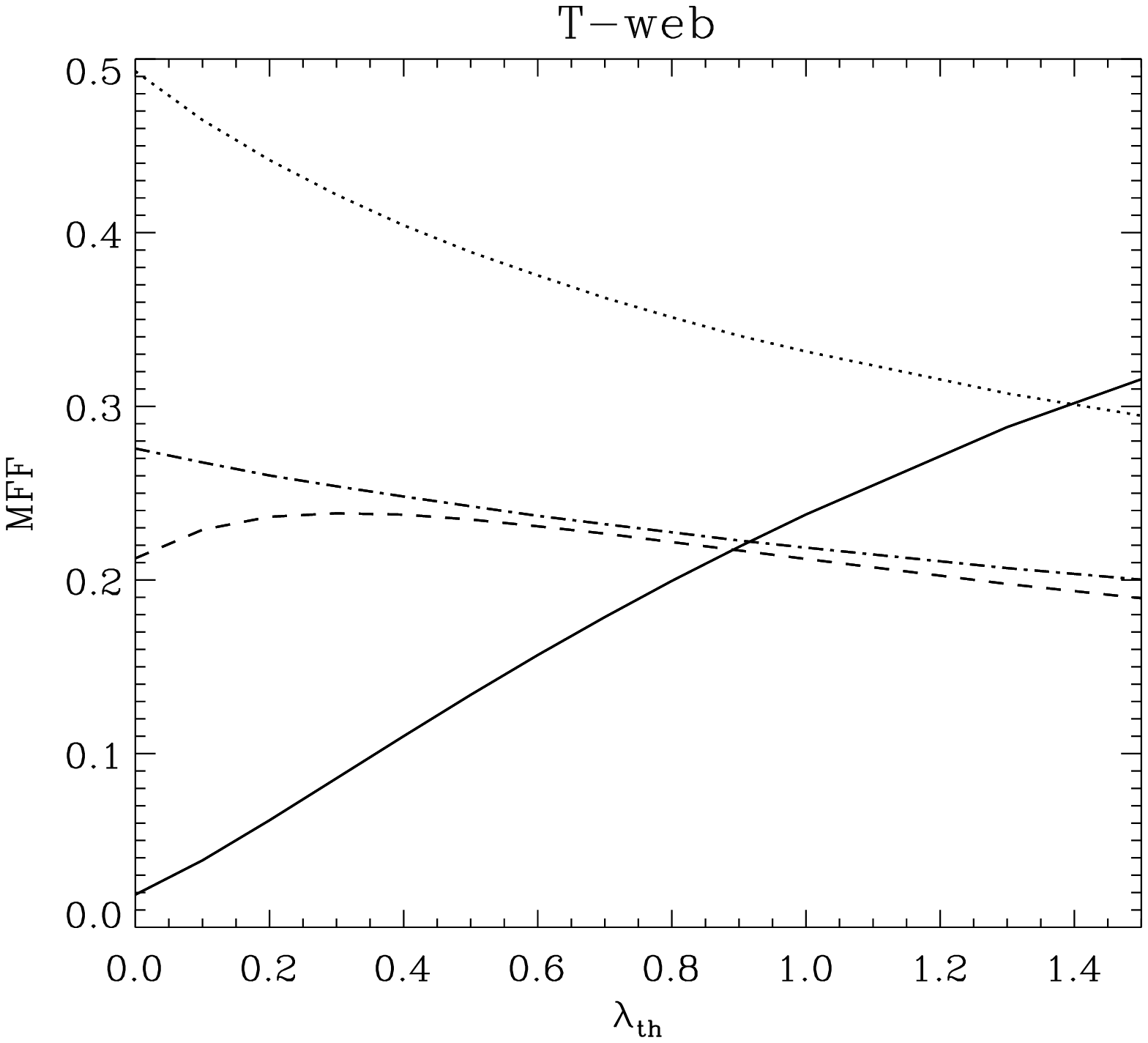}
  \includegraphics[width=0.49\textwidth]{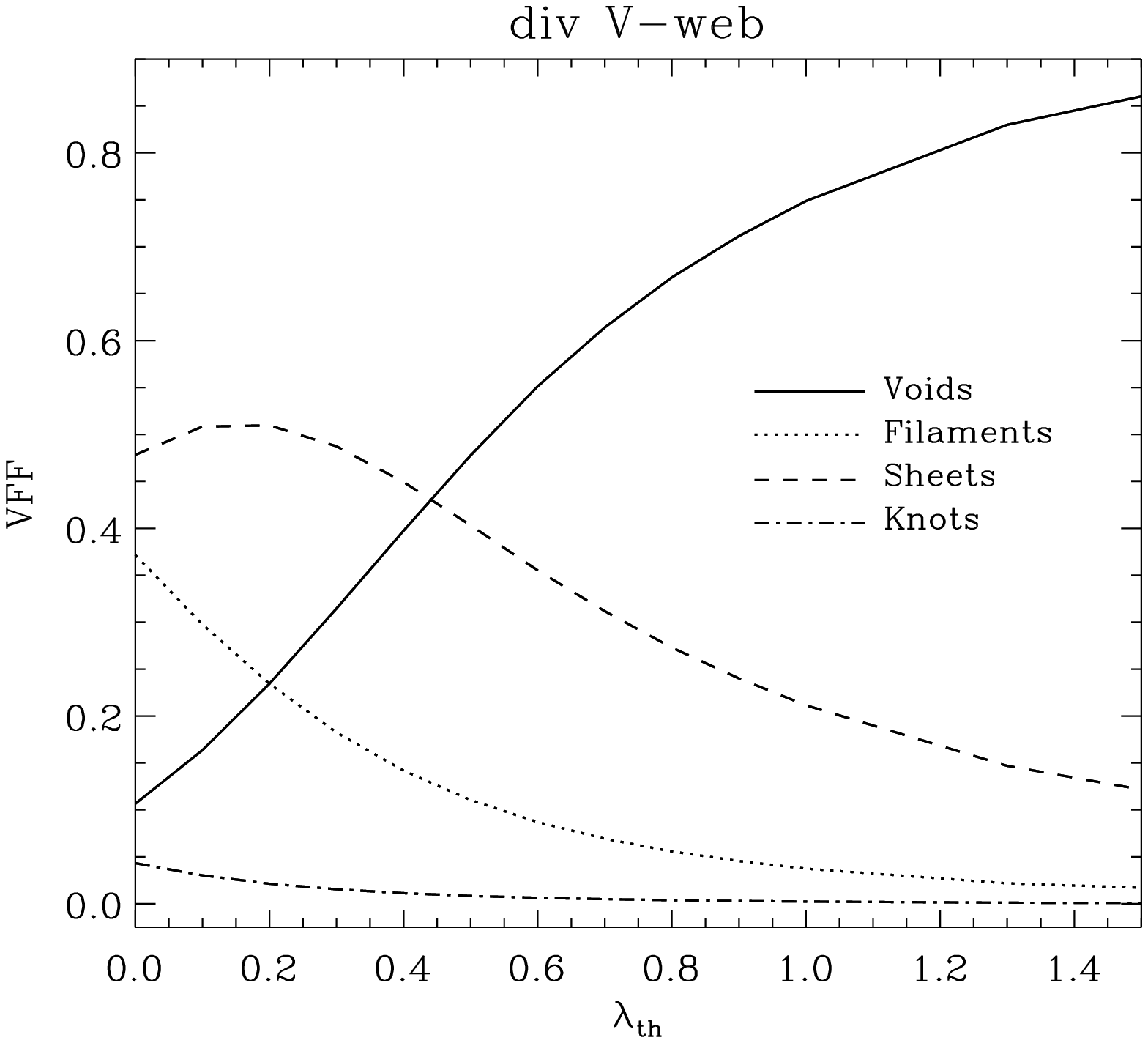}\includegraphics[width=0.49\textwidth]{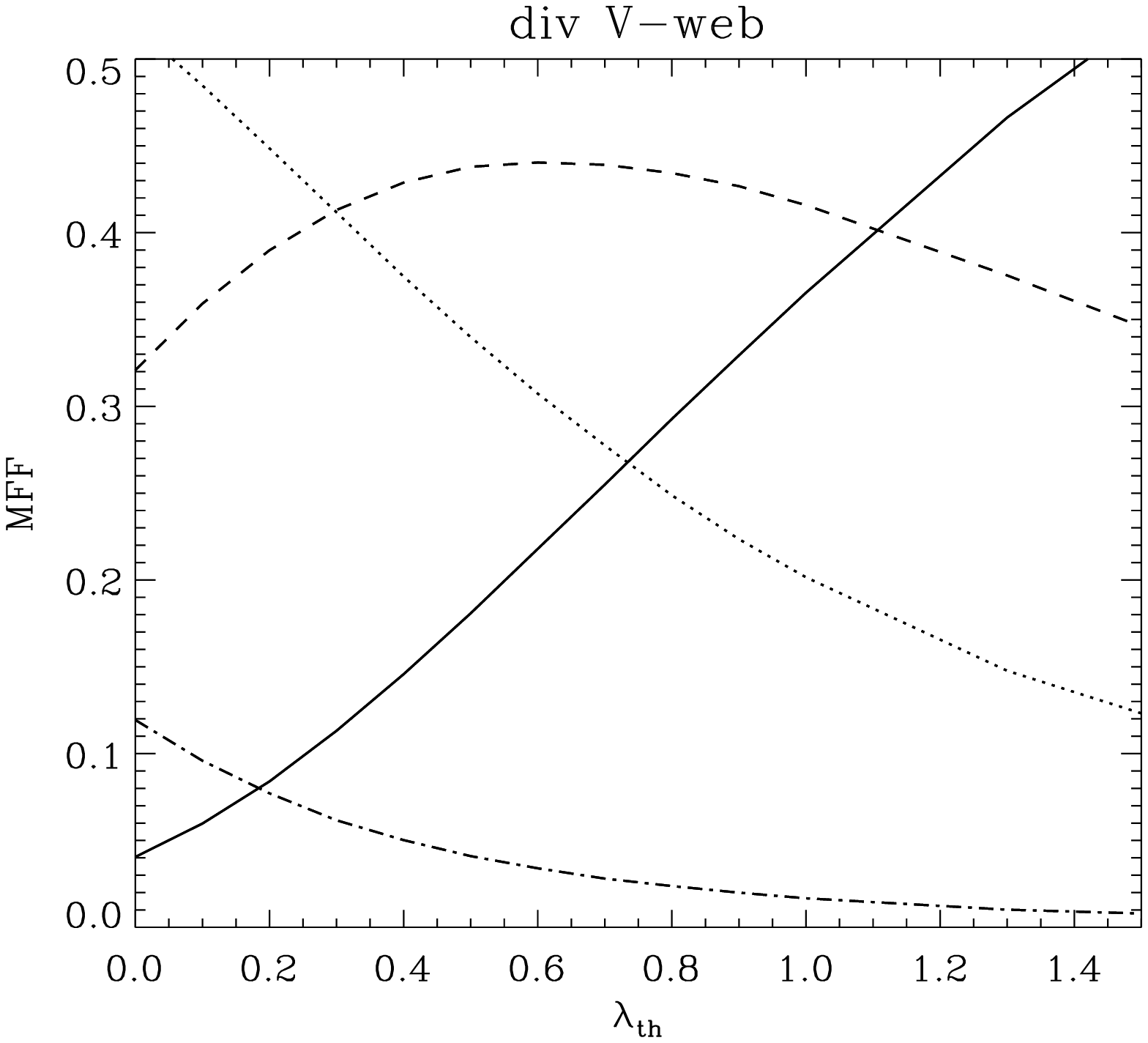}  
  \caption{VFF (left panel) and MFF (right panel) of our best correlated LU 
  reconstruction in real space as a function of threshold eigenvalue for the T- and ${\rm div}\,$V-web cosmic web classification 
  methods. Different environments are indicated.}
\label{fig:Tweb_lth}
\end{figure*}

Interestingly, the halo-derived matter density shape is similar to that obtained when all matter 
is taken into account. In particular, we are able to reproduce the observed increase in matter 
density as the scale radius decreases, an effect that is particularly prominent in our Local Volume 
(LV, i.e. $r\lesssim20\,\Mpch$). Additionally, the simulated halo matter density shows similar bumps 
to observations at distances of about $10-30\,\Msunh$ which are mainly owing to the presence 
of the Local Supercluster. 
At scales below $10\,\Mpch$, the mismatch between data and the matter density of haloes is likely 
owing to a combination of the limited resolution of our simulation and the uncertainties 
present in the selection function used in the reconstruction.

These results demonstrate that the `missing DM' in the LU can be simply interpreted as being 
located outside massive haloes, i.e. forming part of the cosmic web `field' in agreement with the 
$\Lambda$CDM expectations. This interpretation is strengthened by the results of 
Section~\ref{sec:cweb_stats}. There it will be shown that, irrespective of the cosmic web classification used, 
most of the mass is located outside the densest regions.

\begin{figure*}
  \begin{tabular}{cc}
\hspace{1.3cm}
  \includegraphics[width=0.49\textwidth]{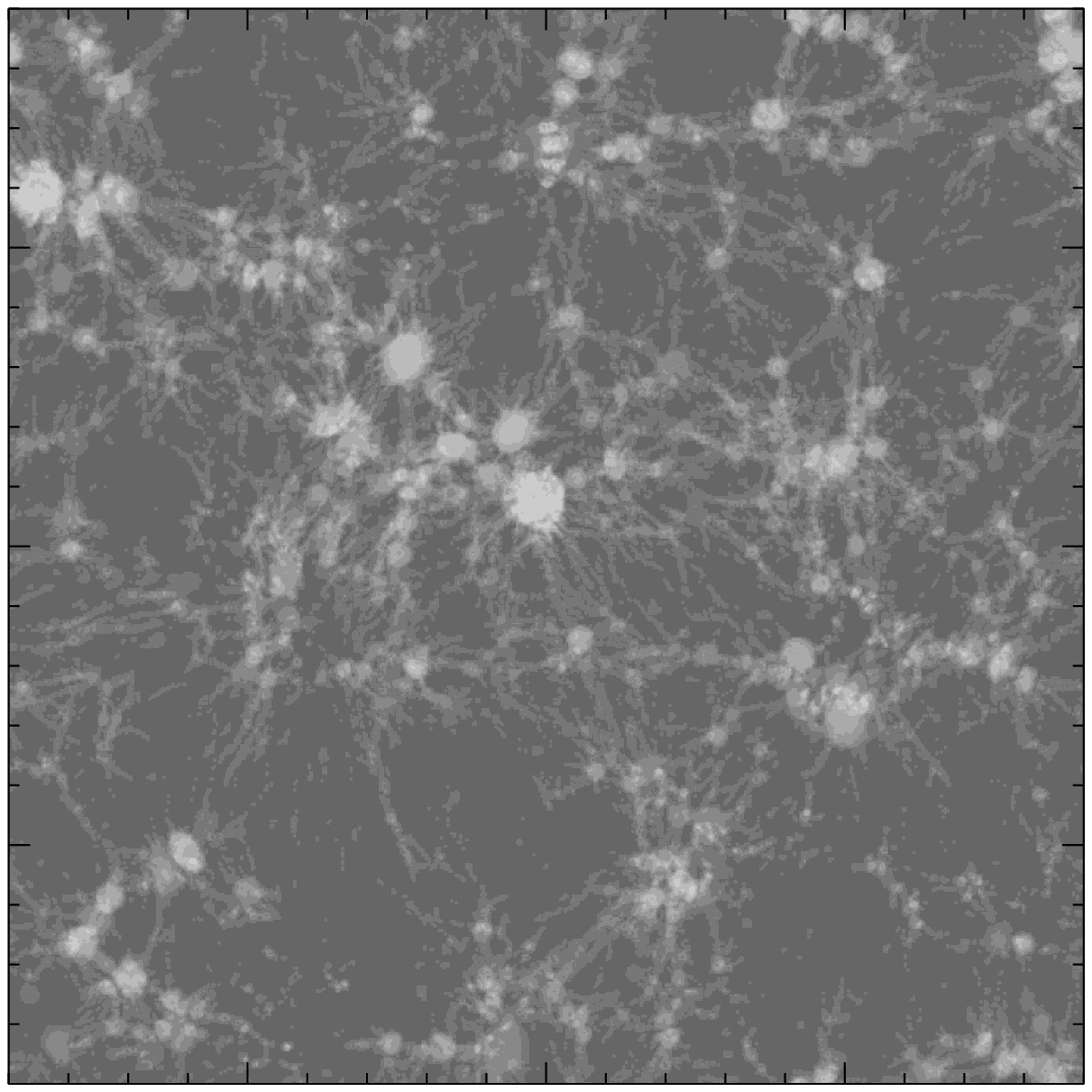}
  \put(-270,135){\rotatebox[]{90}{SGY [$h^{-1}$ Mpc]}}
  \put(-249,185){$50$}
  \put(-245,130){$0$}
  \put(-257,74){$-50$}
  \hspace{-1.65cm}
  \includegraphics[width=0.49\textwidth]{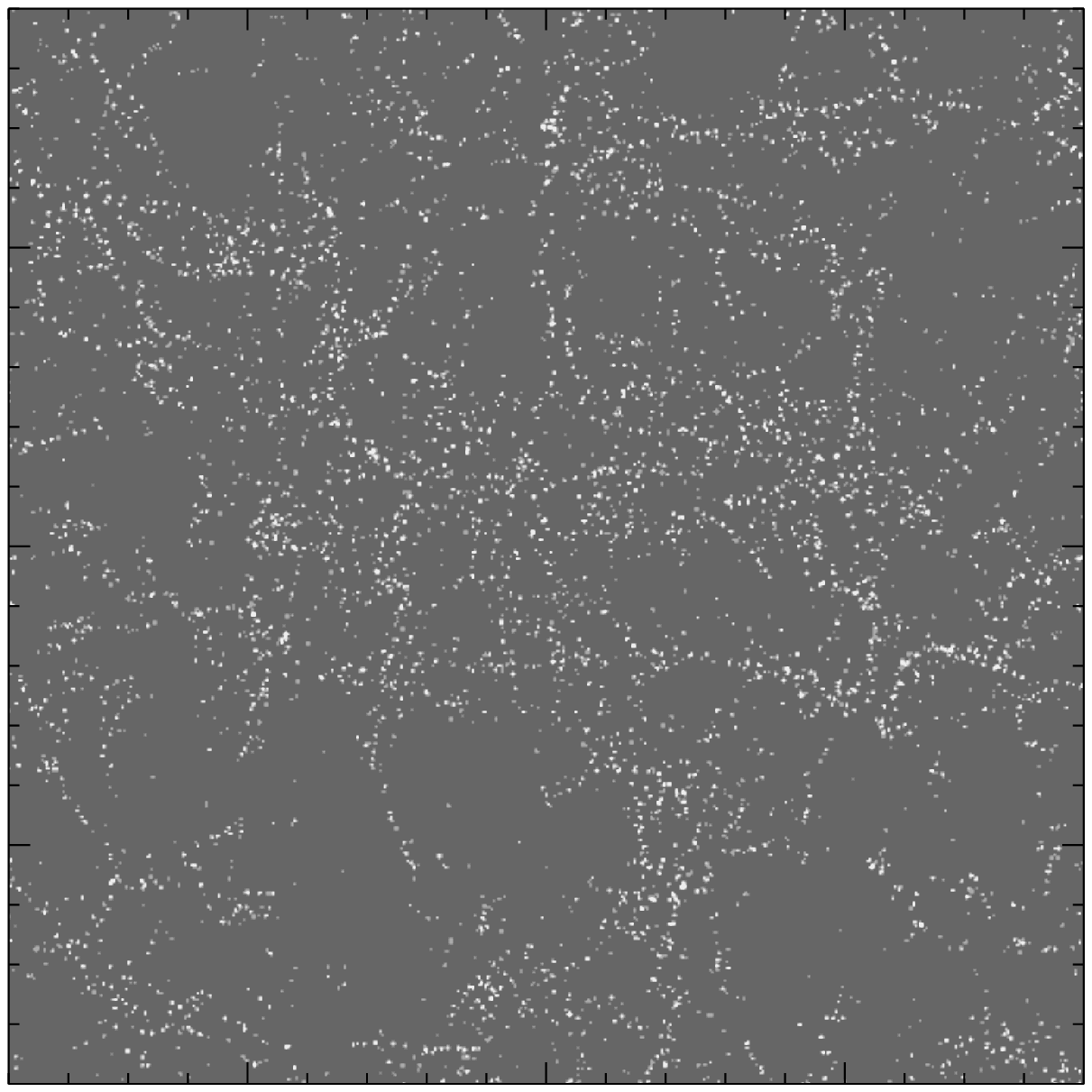}
  \put(-12,135){\rotatebox[]{-90}{SGY [$h^{-1}$ Mpc]}}
  \put(-32,185){$50$}
  \put(-32,130){$0$}
  \put(-32,74){$-50$}
  \vspace{-1.65cm}
  \\
\hspace{1.3cm}
  \includegraphics[width=0.49\textwidth]{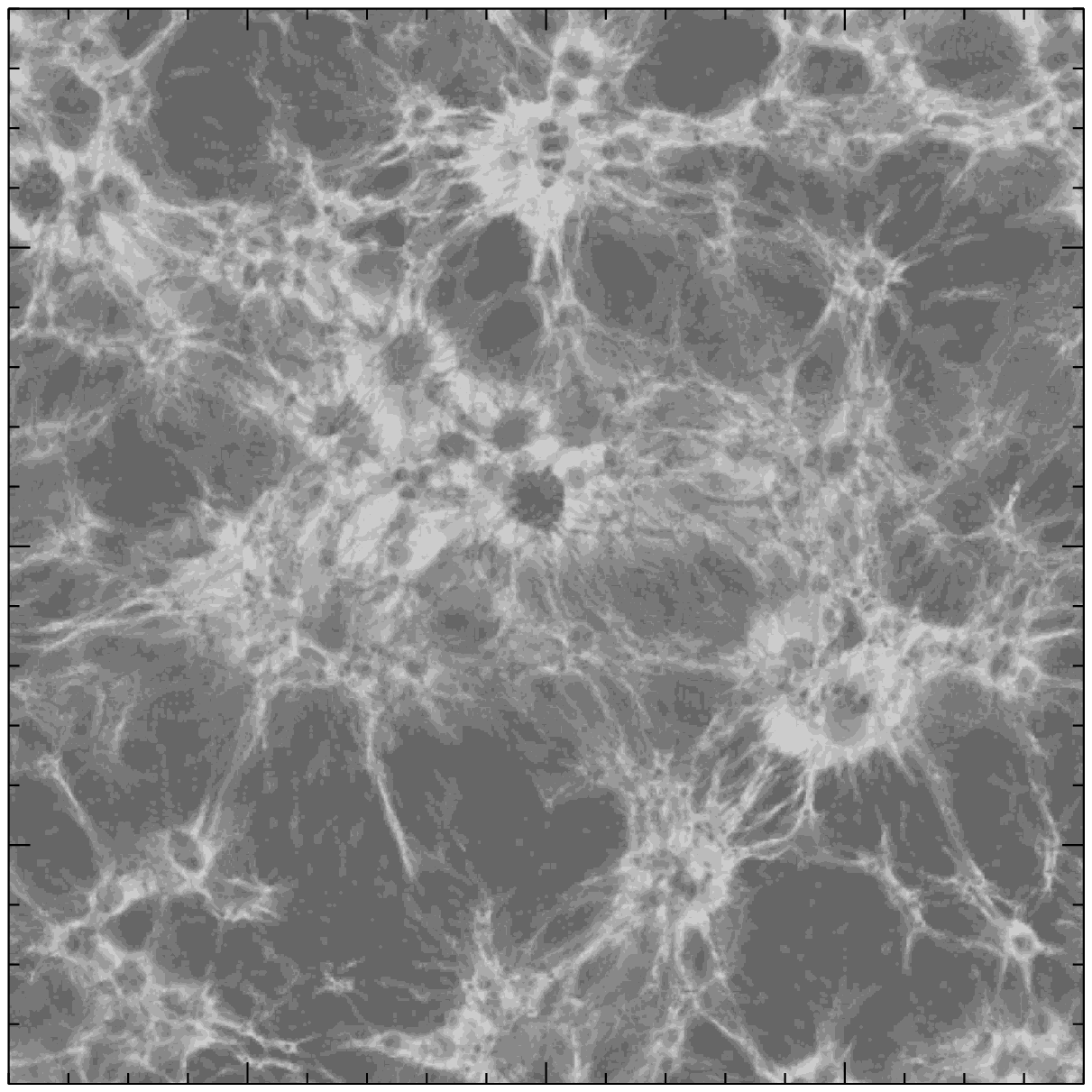}
  \put(-270,135){\rotatebox[]{90}{SGY [$h^{-1}$ Mpc]}}
  \put(-249,185){$50$}
  \put(-245,130){$0$}
  \put(-257,74){$-50$}
  \put(-165,5){\rotatebox[]{0}{SGX [$h^{-1}$ Mpc]}}
  \put(-203,20){$-50$}
  \put(-138,20){$0$}
  \put(-85,20){$50$}
  \hspace{-1.65cm}
  \includegraphics[width=0.49\textwidth]{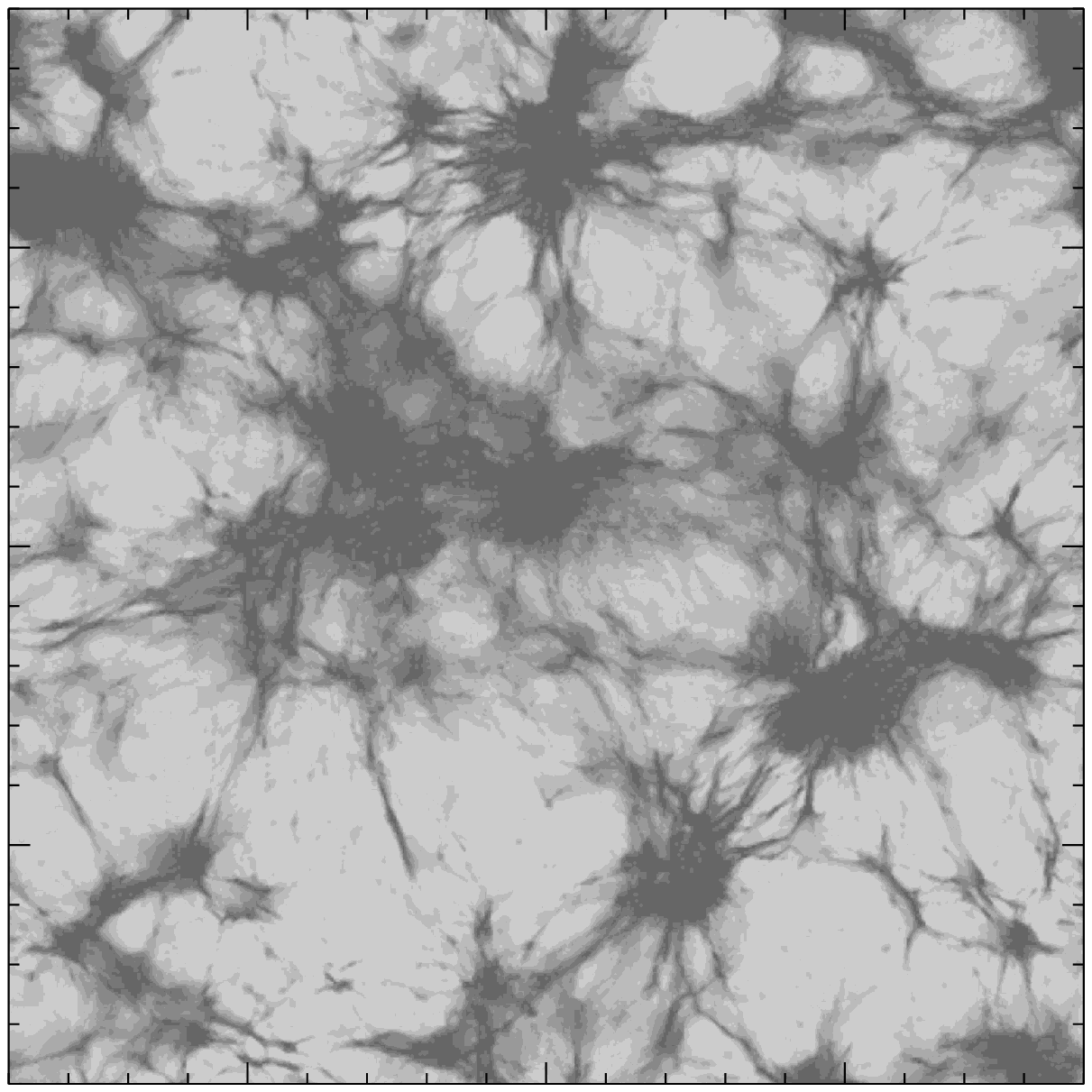}
  \put(-12,135){\rotatebox[]{-90}{SGY [$h^{-1}$ Mpc]}}
  \put(-32,185){$50$}
  \put(-32,130){$0$}
  \put(-32,74){$-50$}
  \put(-165,5){\rotatebox[]{0}{SGX [$h^{-1}$ Mpc]}}
  \put(-203,20){$-50$}
  \put(-138,20){$0$}
  \put(-85,20){$50$}
\end{tabular}
  \caption{Cosmic web density field of our best-correlated LU reconstruction in real space as obtained 
           with the T-web method using 
           $\lambda_{\rm th}=0.9$ for filaments (upper left), knots (upper right), sheets (lower left) 
           and voids (lower right) in a slice of $7\,\Mpch$. The different web structures 
           are shown in light colours.
           The resulting VFFs of the web elements are about 7.6\%, 0.6\%, 20.4\% and 71.4\% 
           respectively (see Section~\ref{sec:cweb_stats}). Units are in supergalactic coordinates.}
\label{fig:cosmic_web_LU}
\end{figure*}

\section{The cosmic web of the Local Universe}
\label{sec:cosweb}

To carry out the cosmic web characterisation of the LU we apply to our reconstruction 
the tidal field tensor classification proposed by \cite{Hahn07a}. 
In addition, we also explore the possibility of using the reconstructed 
nonlinear velocity field to perform the classification. Throughout this paper, 
we will refer to these approaches as the `T-web' and `${\rm div}\,$V-web' respectively. 
These methods are implemented within the \textsc{classic} software package, 
which will be presented in a forthcoming publication (Kitaura \& Nuza, in preparation).

\subsection{The `T-web' method:}

We study the dynamics of matter by computing the eigenvalues $\lambda_i$ $(i=1,2,3)$ of the 
tidal field tensor 

\begin{equation}
  T_{ij}\equiv\frac{\partial^2\phi_{\rm rec}}{\partial x_i \partial x_j},
  \label{eq:T_ij}
\end{equation}

\noindent where $\phi_{\rm rec}$ is the gravitational potential of the reconstructed 
nonlinear density field. By analysing the signature of the eigenvector 
$\vec{\lambda}=(\lambda_1,\lambda_2,\lambda_3)$ at a given spatial point 
it is then possible to distinguish between different dynamical behaviours. 
For instance, collapsing (expanding) structures in all spatial dimensions will 
be classified as halo-like (void-like) regions. In a similar way, 
elongated-like (sheet-like) structures can be identified in the case 
of a 1-dimensional (2-dimensional) expansion. In what follows, we will 
designate to these four cases 
as {\it knots}, {\it voids}, {\it filaments} and {\it sheets} respectively.

However, not all positive (negative) eigenvalues will give rise to an actual 
collapse (expansion) along the corresponding axis in the inmediate future. 
Therefore, to achieve a more realistic correspondence between the tidal field 
tensor dynamical classification and the actual cosmic web we followed 
the approach of \citet{Forero09}: we classify structures according to 
the number of eigenvalues ($N_{\lambda}$) above a certain threshold 
($\lambda_{\rm th}$) which, in general, will be nonzero. 
As a consequence, cells with $N_{\lambda}=3,2,1,0$ will then be classified as 
knots, filaments, sheets and voids, respectively. 

\subsection{The `div\,V-web' method:}

As an alternative way of performing the cosmic web classification, we use the information 
of the reconstructed nonlinear velocities alone. By taking the divergence of the velocity field we 
can infer the expected density distribution under the linear approximation. 
Specifically, the linear density field is obtained by computing

\begin{equation}
  % -Div*v/(f*H*a) = linear_overdensity
  \delta_{\rm lin}=-\frac{\nabla\cdot{\bf{v}_{\rm rec}}}{f H a},
  \label{eq:delta_lin}
\end{equation}

\noindent where $\bf{v}_{\rm rec}$ is the reconstructed nonlinear velocity field, $f$ is the 
logarithmic derivative of the linear growth factor, $H$ is the Hubble constant and $a$ is 
the expansion factor. 
Then, for computing the web, we use the corresponding linear potential as an input of 
the tidal field tensor 
classification method discussed above.      

It is worth noting that in this method -- as well as in any other classification derived from 
the velocity field -- caution must be taken when considering knots, as shell crossing is known 
to dramatically affect the gridded peculiar velocity field (see e.g., \citealt{Hoffman12,Hahn14}).

\subsection{Cosmic web statistics}
\label{sec:cweb_stats}

Fig.~\ref{fig:Tweb_lth} shows the volume and mass filling fractions
(VFFs and MFFs respectively) of the simulated LU as a function of threshold eigenvalue
for the best-correlated LU reconstruction of \citet{Hess13} in real space as obtained
from the T-web (upper panels) and ${\rm div}\,$V-web (lower panels) classification methods. 
As expected, the relative fraction of occupied volume and mass of the different structures 
vary as the threshold eigenvalue is increased. Interestingly, 
the VFFs for the two classification methods studied show somewhat similar values 
for a given eigenvalue threshold. Nevertheless, differences can be large for $\lambda_{\rm th}\approx0$. 
For larger thresholds typical differences are of the 
order of $10-20\%$. This is not the case, however, for the MFFs. Despite 
showing the same trends, the amount of mass in each environment can be significantly 
different between the two methods. This is because the ${\rm div}\,$V-web 
classification linearises the density field thus removing material from the densest regions. 
As a result, the amount of mass in the classified knots and filaments is dramatically 
reduced, whereas the opposite occurs in sheets and voids. This is not
surprising as the two classification methods considered probe different
density regimes. Therefore, we do not expect a match between the corresponding
statistics. In what follows, however, we plan to apply both 
cosmic web `finders' to our reconstructed LU to assess the robustness 
of our conclusions.

\subsubsection{Effective threshold}

To choose the effective threshold eigenvalue defining the cosmic web, we rely
on results of alternative classification methods. In particular, our aim is 
to reproduce the VFF of voids in a $\Lambda$CDM universe which is found to be
considerably larger than about $17\%$, i.e. the value obtained in the
pioneering classification of \cite{Hahn07a}. In general, such alternative
methods predict a void VFF in the range $70-90\%$. Techniques based on
phase-space tesselations of the particle distribution have been used to
classify voids as regions free of shell crossings
\citep{Abel12,Falck12,Shandarin12}. In this way, \cite{Shandarin12} have
shown that the volume fraction occupied by voids could be as high as
$93\%$. However, as noted by \cite{Abel12}, these estimates must be taken with
caution as they are strongly dependent on resolution. Other works, based on
dynamical and/or kinematical approaches, present somewhat smaller values. For
instance, by analysing the velocity shear tensor, \cite{Hoffman12}
presents a cosmic web classification resulting in a $69\%$ of voids. More
recently, \cite{Cautun13} performed a multiscale method, which includes a
density filtering in logarithmic space, predicting a void VFF of $78\%$. 
This diversity in the VFFs hints at the complexity in the classification of the
cosmic web. Therefore, in this work, we adopt a simple approach: we choose a
threshold eigenvalue of $\lambda_{\rm th}=0.9$ to reproduce a void VFF within
the range of estimates mentioned above in our unconstrained simulations. The same
parameter is then consistently used in our reconstructions. 
As a result of this choice the remaining cosmic web structures 
(i.e., knots, filaments and sheets) will be uniquely determined as can be 
inferred from Fig.~\ref{fig:Tweb_lth}. 
We have checked that adopting other values -- that will modify the
corresponding void VFF -- does not change the validity of our
conclusions.

\begin{figure*}
  \includegraphics[width=0.49\textwidth]{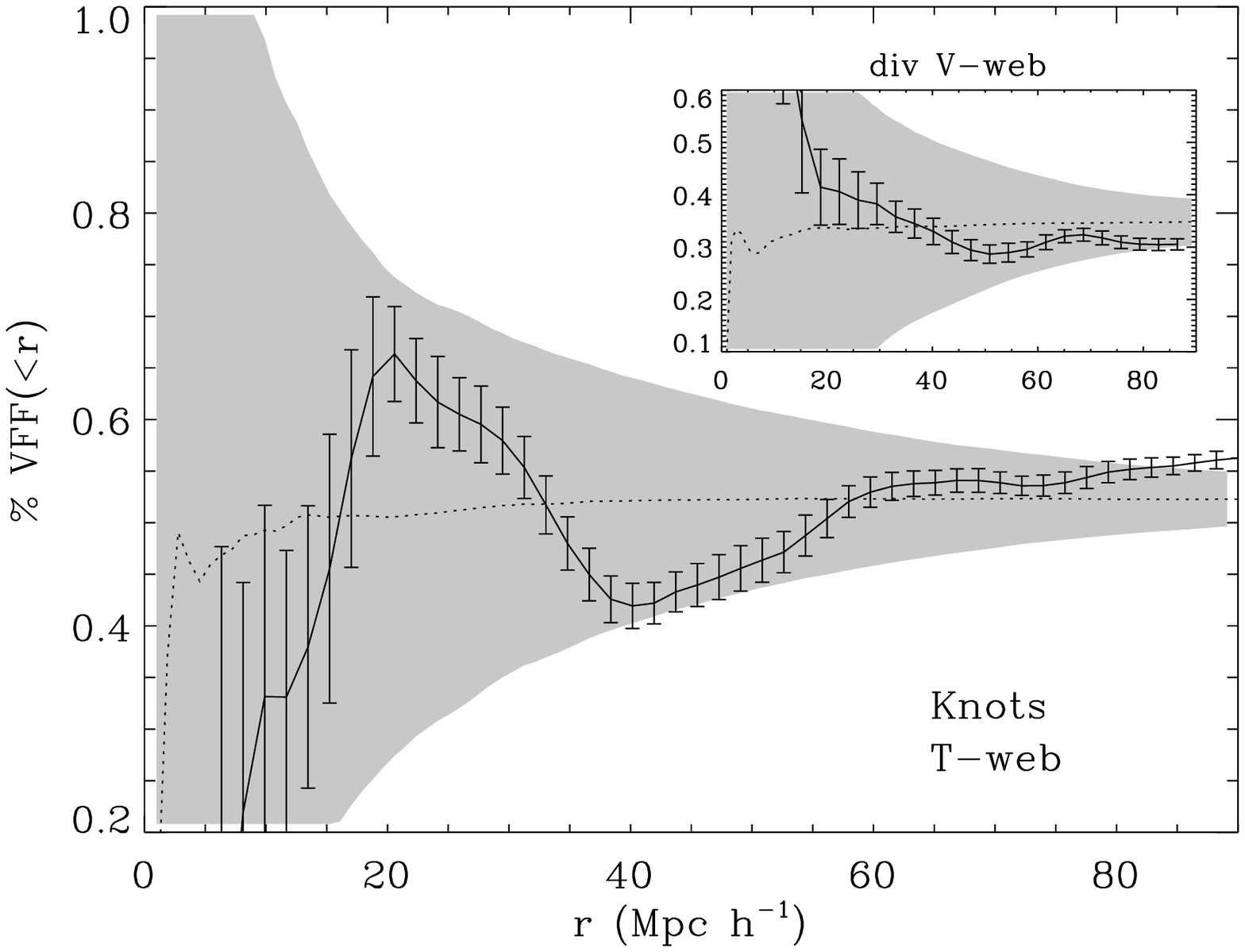}\hspace{1mm}\includegraphics[width=0.49\textwidth]{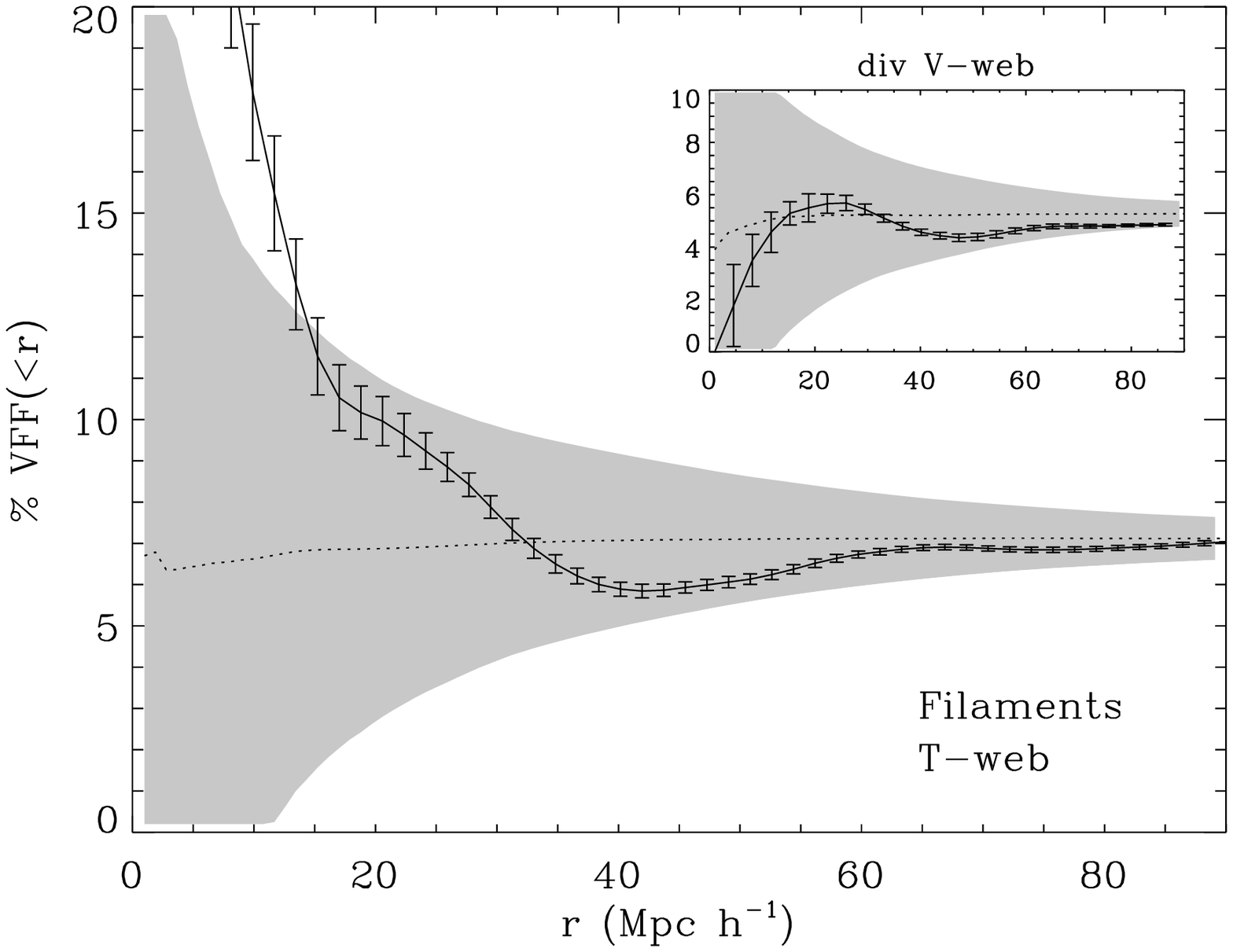}
  \includegraphics[width=0.49\textwidth]{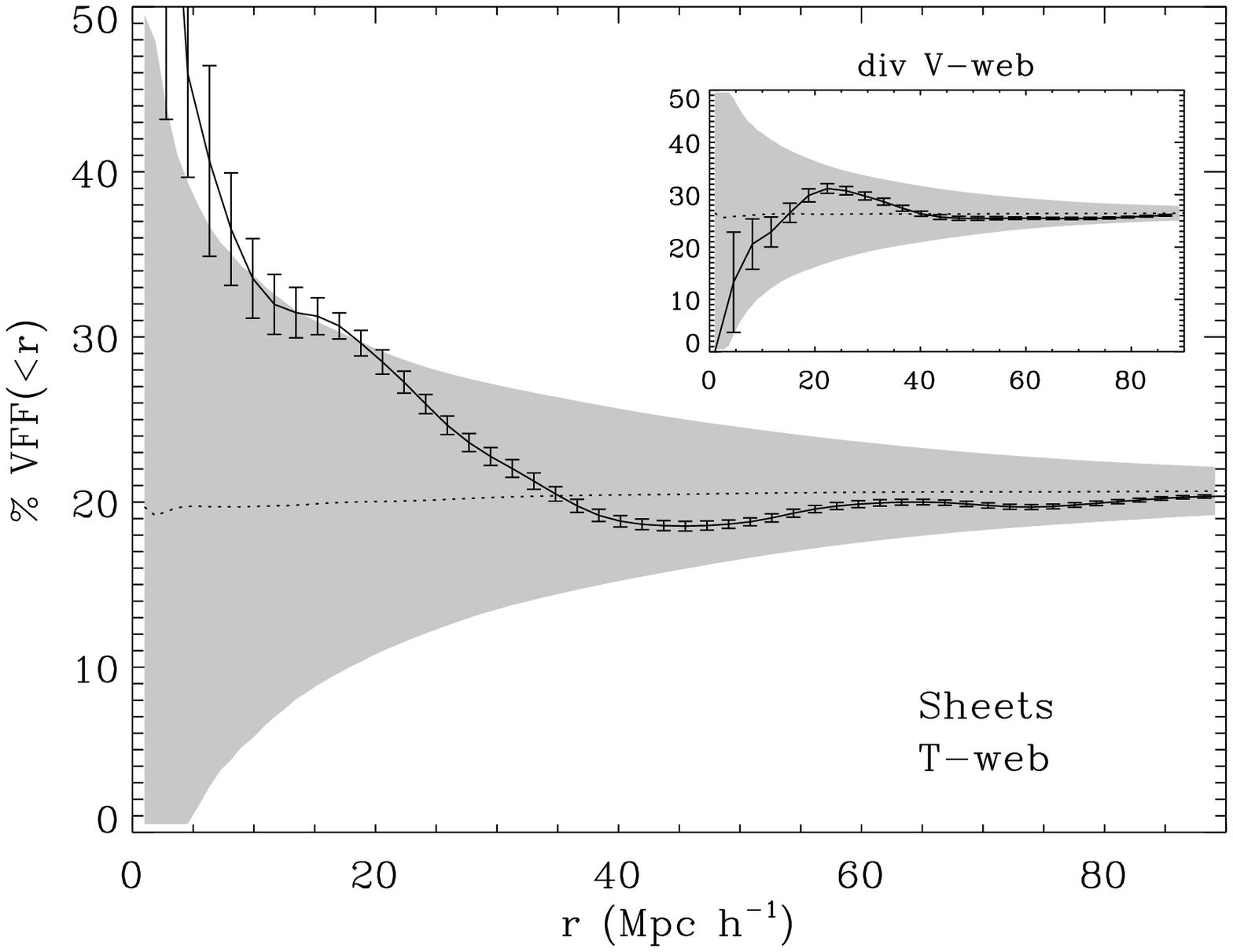}\hspace{1mm}\includegraphics[width=0.49\textwidth]{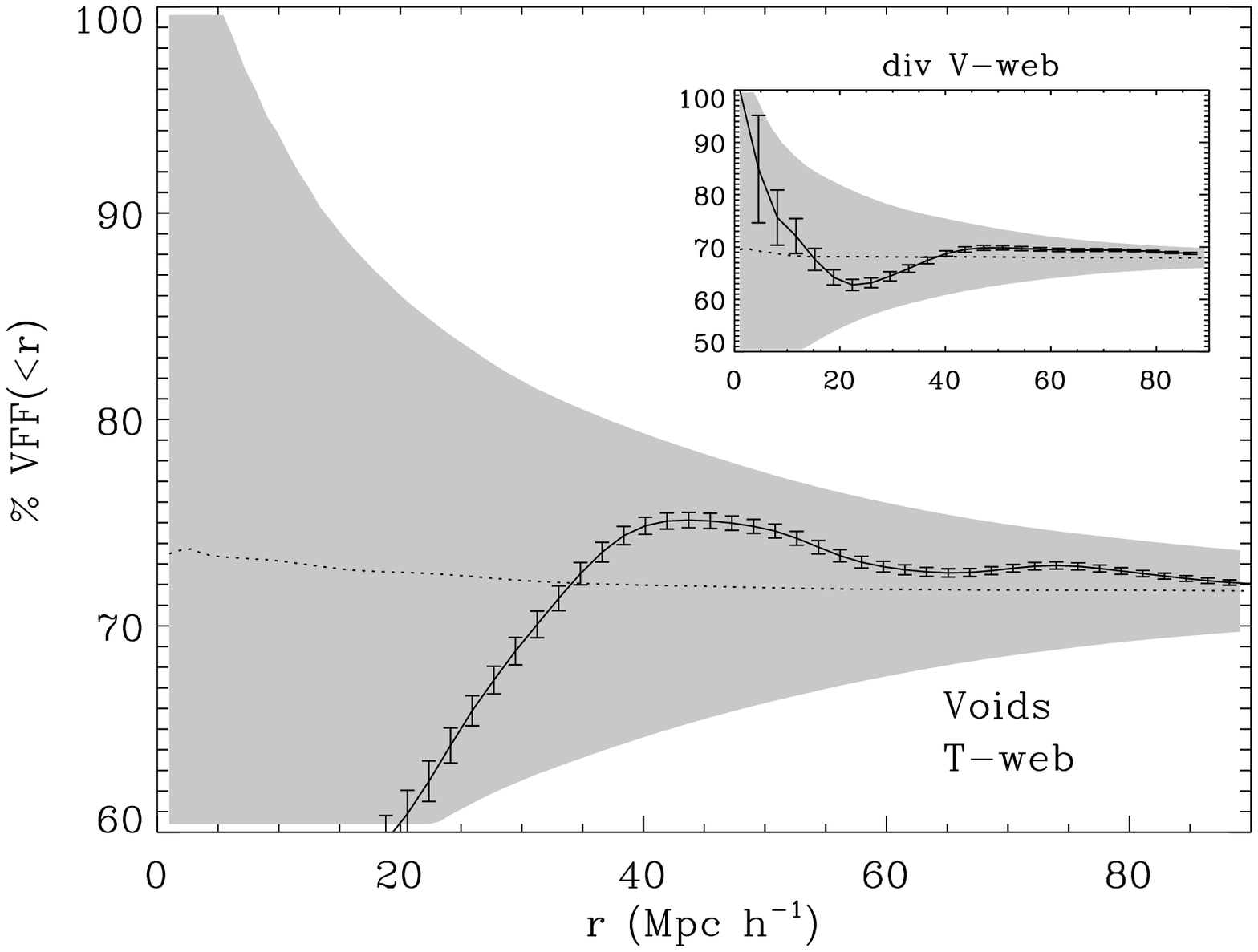}
  \caption{Cumulative VFF of the best-correlated reconstructed LU as a function of 
  distance from the centre of the box for different environments. The cosmic web has been computed 
  using the T-web method for a threshold of $\lambda_{\rm th}=0.9$. The inset plots show the results 
  obtained using the {\rm div}\,V-web classification with $\lambda_{\rm th}=0.8$. 
  The error bars show the standard deviation of our ensemble of 25 reconstructions. 
  The shaded regions represent the 1$\sigma$ cosmic variance fluctuations as obtained from unconstrained $N$-body simulations by 
  placing 1000 `observers' at different locations within the box, whereas the dotted line stands 
  for the mean value.} 
\label{fig:VFF_vs_r_lth}
\end{figure*}

\begin{figure*}
  \includegraphics[width=0.49\textwidth]{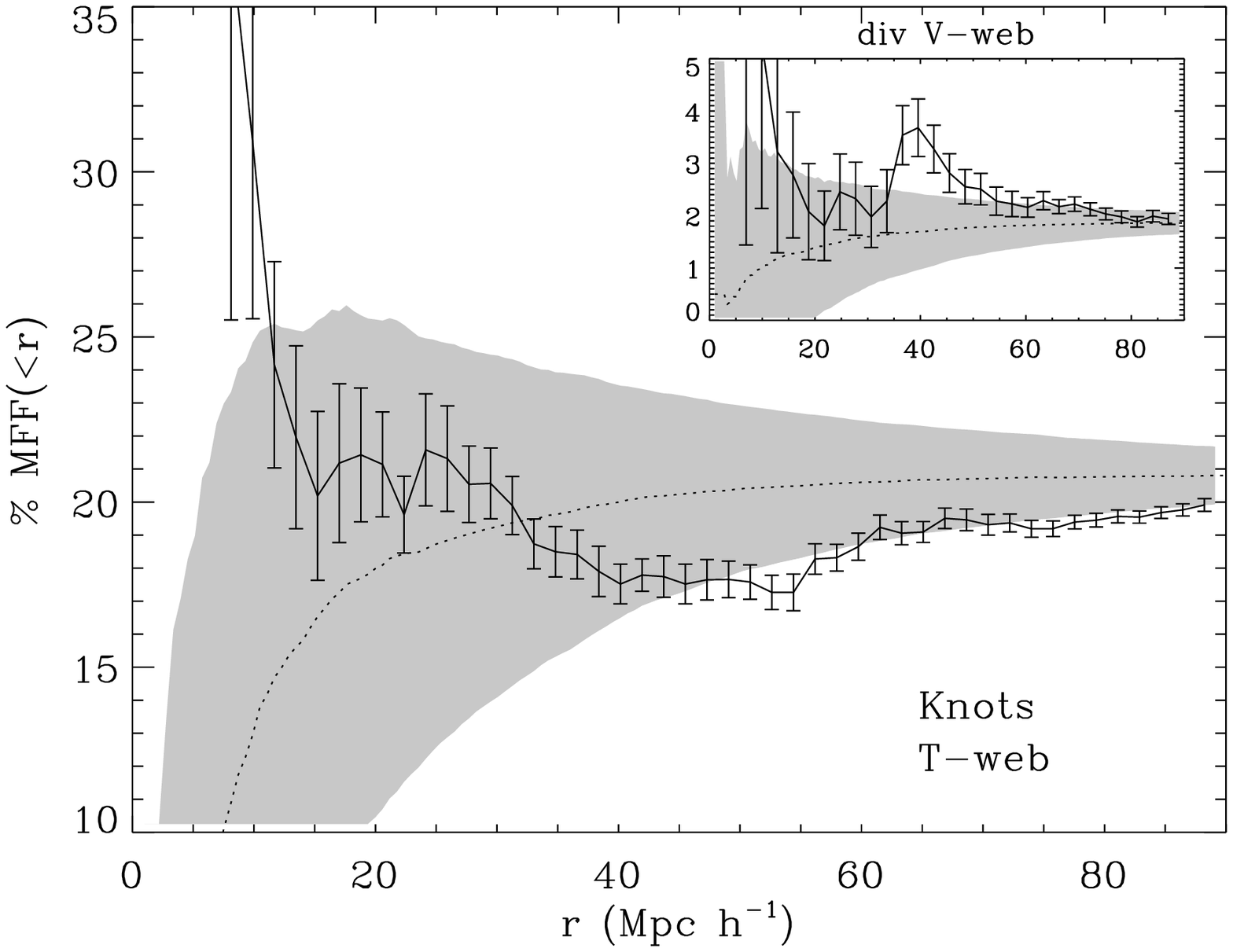}\hspace{1mm}\includegraphics[width=0.49\textwidth]{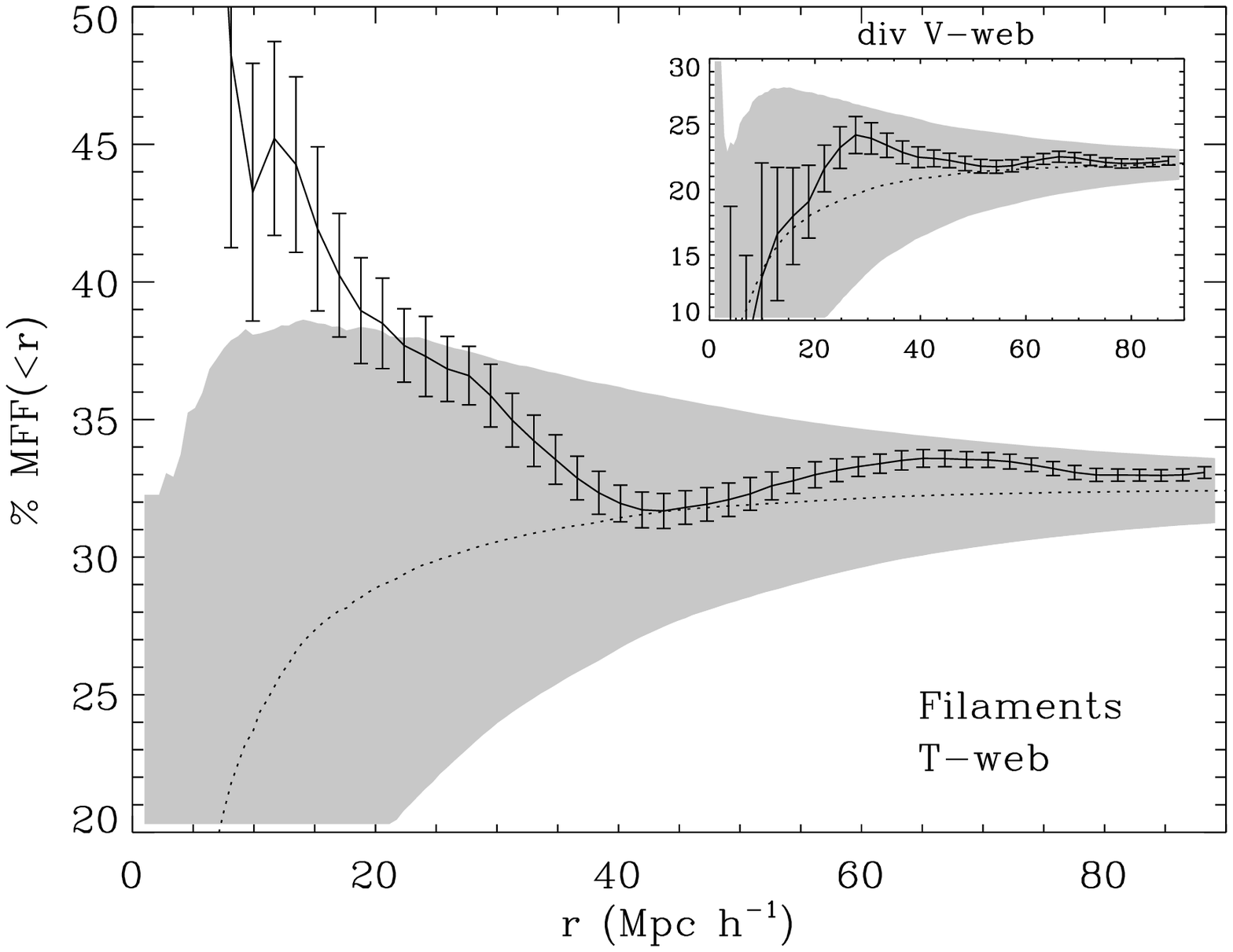}
  \includegraphics[width=0.49\textwidth]{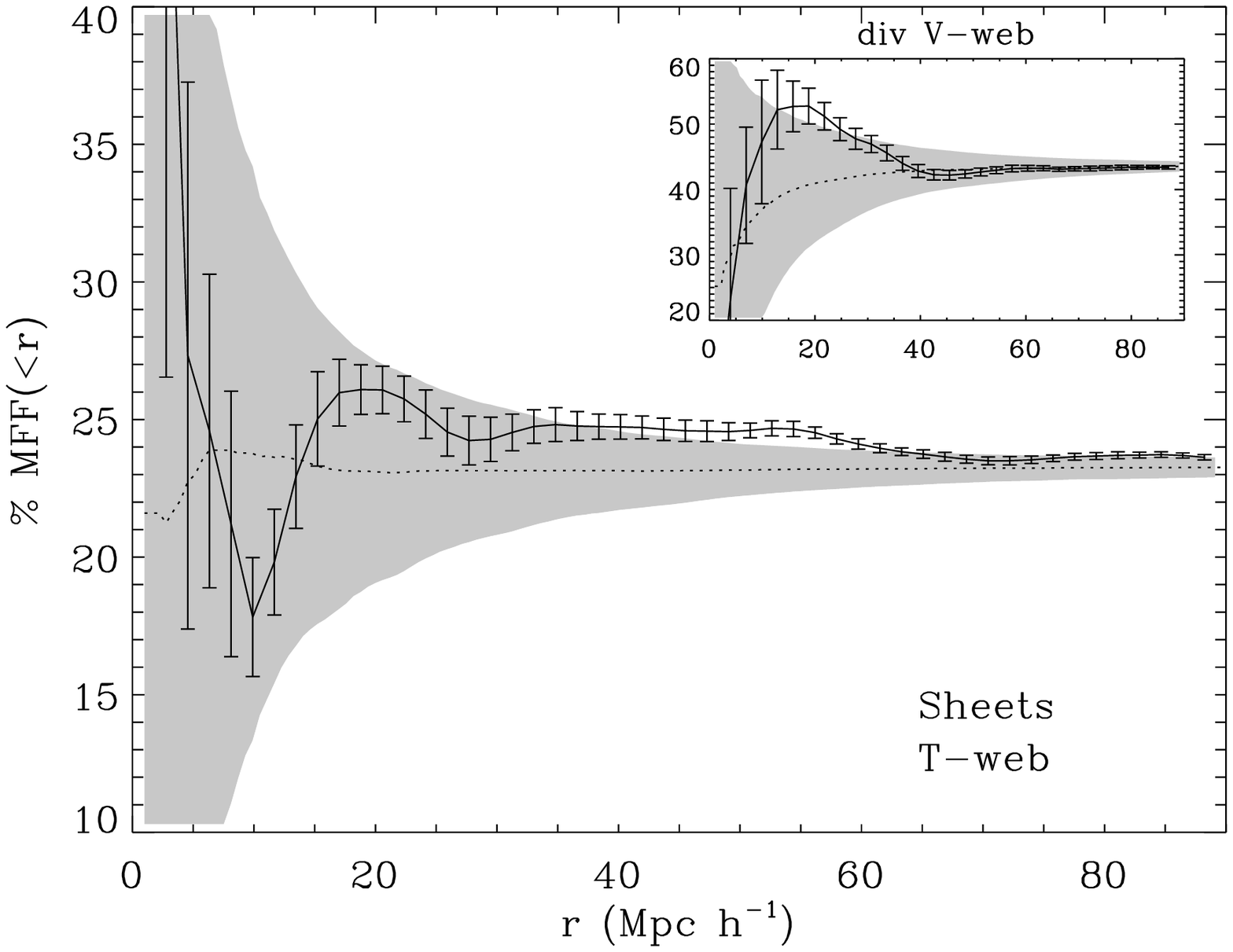}\hspace{1mm}\includegraphics[width=0.49\textwidth]{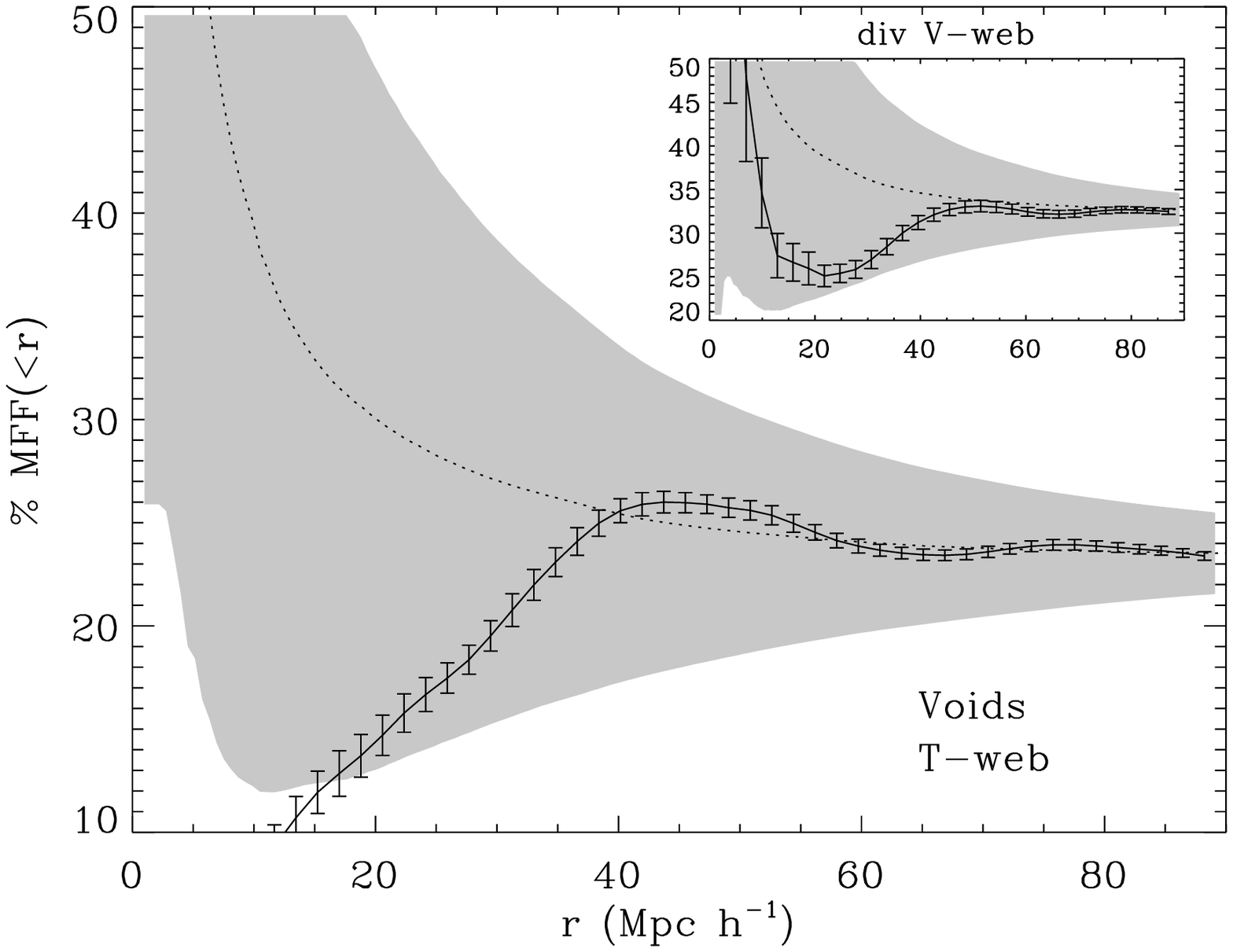}
  \caption{Idem as Fig.~\ref{fig:VFF_vs_r_lth} for the cumulative MFF.}
\label{fig:MFF_vs_r_lth}
\end{figure*}

\subsubsection{Reconstructed filling fractions}

After applying the T-web classification with $\lambda_{\rm th}=0.9$ 
to our best-correlated $N$-body reconstruction in configuration space, we obtain 
VFFs of about $71.4\%$, $20.4\%$, $7.6\%$, $0.6\%$ for voids, sheets, filaments and knots
respectively. Similarly, the corresponding MFFs for these structures are of
about $21.9\%$, $21.7\%$, $34.1\%$, $22.3\%$. We also found that, given the 
resolution adopted here, the resulting statistics are not significantly affected 
if the cosmic web classification is performed in redshift space. 
Fig.~\ref{fig:cosmic_web_LU} shows a projection of the different cosmic web environments 
in the LU as determined by the T-web classification using $\lambda_{\rm th}=0.9$ 
for our best-correlated reconstruction in real space. 
These plots show a slice containing the supergalactic plane 
within a box of $180\,\Mpch$ on a side that is centred in the observer's position. 

To compare the two classification methods
considered here we use, for the ${\rm div}\,$V-web method, an eigenvalue
threshold of $\lambda_{\rm th}=0.8$, aiming at approximately obtaining the
same VFF of voids as with the T-web. In this case, the resulting VFFs and
MFFs are of about $66.7\%$, $27.3\%$, $5.6\%$, $0.38\%$ and $29.3\%$,
$43.5\%$, $24.9\%$, $2.4\%$ for voids, sheets, filaments and knots
respectively for the same LU reconstruction. The small MFFs obtained 
for knots in this case clearly indicates that the ${\rm div}\,$V-web classification fails in
characterising the densest regions, as expected.

\section{Cosmic variance in the LU}
\label{sec:cosmic_variance}

We compare the volume and mass statistics of the reconstructed cosmic web with
those of the random simulations to assess the cosmic variance in the LU.
Therefore, the particular method used to characterise the cosmic web is not
relevant, as soon as the same method is applied to both sets of simulations.
This can be seen in Figs.~\ref{fig:VFF_vs_r_lth} and \ref{fig:MFF_vs_r_lth} for
the T- and ${\rm div}\,$V-web classifications. The latter method 
is shown as inset plots to readily compare between the two web `finders' in each case. 
The different panels show the cumulative VFFs (MFFs) of knots 
(upper-left panel), filaments (upper-right panel), sheets (lower-left panel) and voids
(lower-right panel) as a function of distance to the observer, where the solid
lines indicate the results for our best-correlated reconstruction in real space 
and the error bars (standard deviation) have been estimated using our 
ensemble of 25 reconstructions.
To construct these plots we spherically averaged the corresponding
filling fractions for a given scale to simplify the description of the problem. 
The grey shaded regions indicate the 1$\sigma$ cosmic variance level of an
ordinary $\Lambda$CDM universe, built by placing 1000 `observers' at different
locations within our random set of unconstrained $N$-body simulations, whereas
dotted lines indicate the mean value.  We have checked that our results are
marginally affected if, instead, we compute these statistics in redshift
space, given the adopted resolution in our study. 
(Nevertheless, the impact of redshift-space distortions will be further 
considered in Section~\ref{sec:env}.) 
The first conclusion we can extract from these plots is that the LU is
completely consistent with the expectations of $\Lambda$CDM, i.e. the measured
LU statistics are well within the predicted 1$\sigma$ fluctuations of the
concordance model at most of the studied scales and environments (despite of
some departures for distances smaller than $\sim$15$\,\Mpch$ where the
selection function is less constrained).  Irrespective of the classification,
some remarkable features are evident for $r\gtrsim20\,\Mpch$, i.e. as one increases
the distance from the observer. At smaller scales, however, we found some
differences between our two cosmic web methods. Since at these scales the
selection function is not well constrained, in what follows, we will mainly
focus on the results at $r\gtrsim20\,\Mpch$. In particular, for a sphere of
radius $r=20\,\Mpch$ the VFF of our reconstructed web is dominated by voids,
occupying around $60\%$ of the volume, followed by sheets and filaments, that
comprise $30-35\%$ and $5-10\%$ of the available space respectively.  These
values represent a fluctuation of about $1\sigma$ with respect to the expected
mean of a $\Lambda$CDM universe, which is about $70-72\%$ for voids, $20-25\%$
for sheets and $5-7\%$ for filaments at this scale. High density regions, as
characterised by knots, occupy less than $1\%$ of the volume, which is close
to the expectation for an unconstrained universe.  Despite the fact that the
LV is mainly populated by sheets/voids, our immediate vicinity shows a smaller
fraction of void-like regions in comparison to the mean $\Lambda$CDM
expectation. This behaviour is, however, inverted for matter inside spheres
with radius between $30-60\,\Mpch$, where the VFF is above (below)
the mean expectation in the case of voids (sheets, filaments and knots)
reaching a peak (dip) at a distance of about $40\,\Mpch$. This indicates that,
at that particular radius, the fraction of voids is higher than average, thus
contributing with more voids to the corresponding VFF. As expected, for
increasingly larger spheres, 
the VFFs of the different cosmic web structures converge to the global mean value of the 
whole reconstruction. This behaviour is also observed for the MFF case, where we find 
results consistent with $\Lambda$CDM for radii greater than $\sim$$15\,\Mpch$. 
In this case, however, the trends in the cumulative mass fractions present some 
differences between the two web `finders'. In fact, these discrepancies
are expected as the cosmic web resulting from different approaches 
do not completely overlap \citep[e.g.,][]{Cautun13}. This is most noticeable for 
knots owing to the field linearisation carried out in 
the ${\rm div}\,$V-web classification. However, this effect generates differences in other 
cosmic web structures as well.
Therefore, this fact prevents us from making strong conclusions about the cumulative MFF profile 
in these environments. Nevertheless, we can safely 
conclude that, irrespective of the classification, the LU is in agreement 
with the expected $\Lambda$CDM fluctuations as our measured statistics are always 
contained within $1\sigma$ at scales $r\gtrsim15\,\Mpch$. In addition, we can state that, 
when considering a volume with a radius of about $60\,\Mpch$, the LU becomes a fair 
sample thus converging to the mean $\Lambda$CDM expectations.

\section{Environmental dependence of galaxy morphology}
\label{sec:env}

As a first application of our detailed LU reconstructions we use the 2MRS catalogue 
to correlate the position of galaxies with the environmental information provided by the 
cosmic web classifications computed in Section~\ref{sec:cweb_stats}. To avoid 
any gridding effect all fields analysed here will be convolved with a Gaussian 
filter using a length of one cell. Nevertheless, we have checked 
that the resulting trends and significances are slightly affected if no smoothing 
is used. 

We divide the observed galaxy sample in 4 broad 
categories according to the morphological type assigned by the 2MRS 
team, namely: ellipticals (E), lenticulars (S0), spirals (Sp) and irregulars (Irr). 
After selecting objects within the volume of the reconstruction we end up with 
a sample of 21,893 galaxies, out of which 8,367 correspond to ellipticals, 1,323 to lenticulars, 
11,903 to spirals and 300 to irregulars. 
To every galaxy, we assign an environment corresponding to the cell 
where it is placed as determined by a given cosmic web classification method. 
Following \cite{LeeLee08}, we define what we call the `excess probability ratio' as

\begin{equation}
  \eta(\mathcal{T}|\mathcal{E})\equiv\frac{\mathcal{P}(\mathcal{T}|\mathcal{E})}{\mathcal{P}(\mathcal{G}|\mathcal{E})},
\label{eq:delta_excess}
\end{equation}

\noindent where $\mathcal{G}$ represents a random galaxy in the volume, $\mathcal{T}$ is the 
galaxy type, $\mathcal{E}$ is the considered environment (knot, filament, sheet or void) and 
$\mathcal{P}(\mathcal{X}|\mathcal{Y})$ is the conditional probability of $\mathcal{X}$ given 
$\mathcal{Y}$. The interpretation of the excess probability ratio is therefore rather simple: 
if galaxies of a given type displayed some preference to be located in a particular environment 
then the excess probability ratio has to be $\eta(\mathcal{T}|\mathcal{E})>1$, whereas 
$\eta(\mathcal{T}|\mathcal{E})<1$ has to be true in the opposite situation. 
The case $\eta(\mathcal{T}|\mathcal{E})=1$ indicates no distinction from the random distribution.

To estimate the actual conditional probabilities we measure galaxy number counts provided 
by the 2MRS catalogue. Specifically, we approximate their values as 
$\mathcal{P}(\mathcal{T}|\mathcal{E}) \approx N_{\rm g}(\mathcal{T}|\mathcal{E})/N_{\rm g}(\mathcal{T})$ 
and $\mathcal{P}(\mathcal{G}|\mathcal{E}) \approx N_{\rm g}(\mathcal{E})/N_{\rm tot}$, where $N_{\rm g}(\mathcal{Z})$ 
is the number of galaxies satisfying the condition $\mathcal{Z}$ and $N_{\rm tot}$ is the total number of 
galaxies in our sample. Therefore, our estimator for the excess probability presented 
in Eq.~(\ref{eq:delta_excess}) results in

\begin{equation}
  \eta(\mathcal{T}|\mathcal{E})\approx\frac{N_{\rm g}(\mathcal{T}|\mathcal{E})N_{\rm tot}}{N_{\rm g}(\mathcal{T})N_{\rm g}(\mathcal{E})}.	
\label{eq:excess_app}
\end{equation}

\begin{figure*}
  \includegraphics[width=1.0\textwidth]{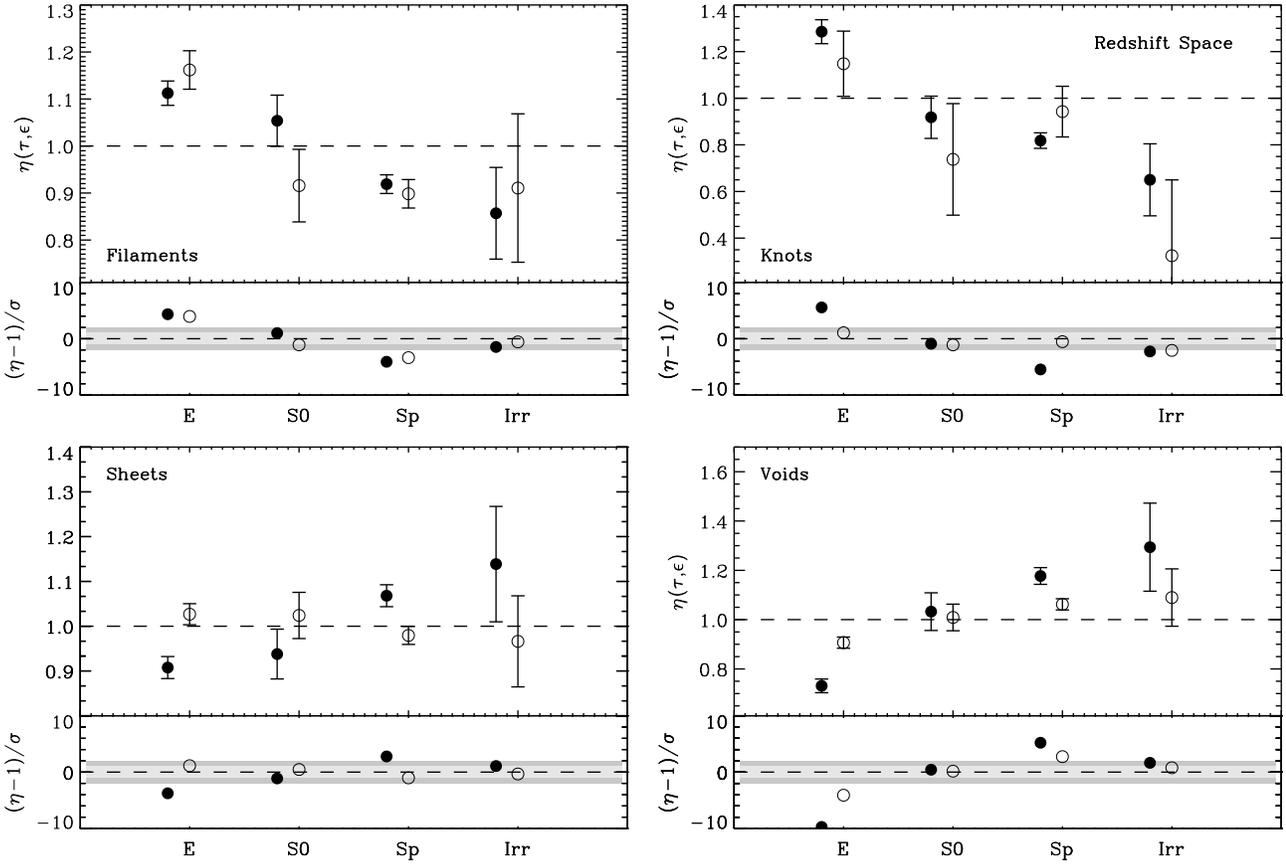}
  \caption{`Excess probability' $\eta(\tau,\epsilon)$ in redshift space 
   for a galaxy of a given morphological type $\tau$ 
   (i.e., E: elliptical; S0: lenticular; Sp: spiral; Irr: irregular) to inhabit a 
   particular environment as indicated in the panels. 
   Results are presented for the T-web (solid circles) and {\rm div}\,V-web (open circles) 
   classification methods using a threshold of $\lambda_{\rm th}=0.9$ and $0.8$ respectively. 
   The error bars are computed assuming Poisson statistics. 
   Also shown is the excess probability with respect to the null signal 
   (as represented by the dashed lines) measured, in each case, 
   in units of the standard deviation, where the light (dark) shaded regions represent
   the 1$\sigma$ (2$\sigma$) levels.}
\label{fig:cweb_gxmorph_z}
\end{figure*}

\begin{figure*}
  \includegraphics[width=1.0\textwidth]{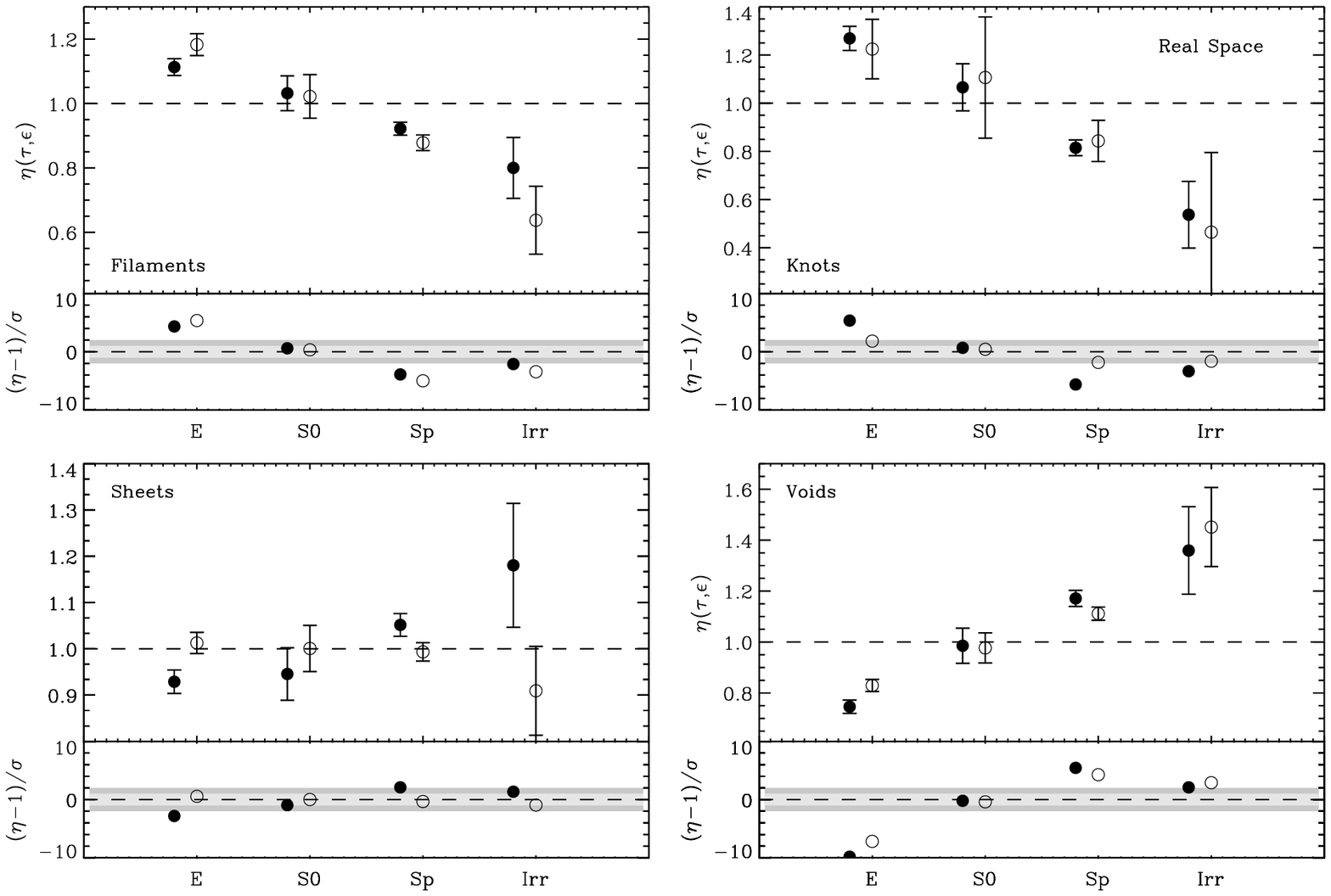}
  \caption{Idem as Fig.~\ref{fig:cweb_gxmorph_z} but in real space. A mapping between the 
  real- and redshift-space reconstruction has been used to assign the most likely real-space 
  position to every 2MRS galaxy (see text).}
\label{fig:cweb_gxmorph_real}
\end{figure*}

The shot noise contribution of the number counts is modelled assuming Poisson
statistics thus allowing us to estimate the magnitude of the associated
errors. 

\subsection{Correlations in redshift space}

Fig.~\ref{fig:cweb_gxmorph_z} shows the resulting excess probabilities
of the different galaxy morphologies within the cosmic web for the
redshift-space LU reconstruction. In this way, we can directly compare the
reconstruction to observations. To compute the excess probability signal we
use both the T-web and {\rm div}\,V-web classification methods.  In the first
place, it can be seen that there is a global tendency for elliptical galaxies
to preferentially reside in knots and in filaments rather than in sheets and
voids. This trend is inverted in the case of spiral and irregular
galaxies. This can be more clearly seen in the T-web case whereas, when
adopting the {\rm div}\,V-web method to define the web, the error bars tend to
increase thus erasing the signal corresponding to spirals.  However, quite
generally, lenticular galaxies display no preference for a particular
environment irrespective of the classification method. These measurements
provide the strength of the tendency to inhabit/avoid a certain
environment. According to the T-web classification, which provides the most
significant detections, it can be seen that ellipticals are most easily found
in knots ($\eta=1.29 \pm 0.05$) than in filaments ($\eta=1.11 \pm 0.03$)
whereas they tend to avoid sheets ($\eta=0.91 \pm 0.02$) less than voids
($\eta=0.73 \pm 0.03$).  On the contrary, spirals are most commonly found in 
voids ($\eta=1.18 \pm 0.03$) than in sheets ($\eta=1.07 \pm 0.02$) whereas
they tend to avoid filaments ($\eta=0.92 \pm 0.02$) less than knots
($\eta=0.82 \pm 0.03$). In general, irregular galaxies follow the same trends
as spirals although, in this case, the error bars are larger as a result of
their smaller number in our sample. For ellipticals, the significance of the
corresponding signals -- according to the T-web ({\rm div}\,V-web)
classification -- result in: $5.5\sigma\,(1.1\sigma)$,
$4.3\sigma\,(3.9\sigma)$, $3.8\sigma\,(1.1\sigma)$ and
$9.7\sigma\,(4.1\sigma)$ for knots, filaments, sheets and voids, respectively.
Similarly, for spirals, we obtain significances of $5.4\sigma\,(0.5\sigma)$,
$4.1\sigma\,(3.3\sigma)$, $2.8\sigma\,(1\sigma)$ and $5.2\sigma\,(2.7\sigma)$
for the corresponding environments.

\subsection{Correlations in configuration space}

As an additional exercise we also computed the probability ratios using the
reconstructed real-space reconstruction as shown in
Fig.~\ref{fig:cweb_gxmorph_real}.  To compare our reconstruction with
observations, we transformed the observed data from redshift to real
space. This is, in general, not a trivial task although, using our
reconstructions, it is possible to establish a mapping between redshift and
configuration space (see Fig.~\ref{fig:DM_overdens}).  Specifically, we use our
set of constrained haloes in real and redshift space to estimate the most
likely real-space position of the corresponding observed 2MRS galaxies. 
In particular, for each galaxy, we search for the closest 
halo in redshift space using our ensemble of cosmological simulations, and assign 
the associated real-space position of the halo.
As can be seen from Fig.~\ref{fig:cweb_gxmorph_real}, the excess 
probability correlations in real space display, in general, the same trends discussed before. 
However, some new features are worth mentioning. In this case, the correlations show a 
smaller scatter, typically decreasing the size of the error bars. 
Moreover, the $\eta$-morphology relations derived using the 
T-web and the ${\rm div}\,$V-web get more consistent with each other. This is mainly owing to the fact 
that working in configuration space reduces the artificial shell crossing caused by redshift-space distortions. 
For ellipticals, in real space, the significance of the 
corresponding signals -- according to the T-web ({\rm div}\,V-web) classification -- are: 
$5.3\sigma\,(1.8\sigma)$, $4.3\sigma\,(5.3\sigma)$, $2.8\sigma\,(0.5\sigma)$ and 
$9.8\sigma\,(7.2\sigma)$ for knots, filaments, sheets and voids, respectively. 
Similarly, for spirals, we obtain significances of $5.6\sigma\,(1.8\sigma)$, $3.9\sigma\,(5\sigma)$, 
$2.1\sigma\,(0.3\sigma)$ and $5.4\sigma\,(4.3\sigma)$ for the corresponding environments. 
Note that, in general, the results for the T-web and {\rm div}\,V-web methods are now 
more similar to each other than in the redshift-space case, as expected.

Interestingly, for a given environment, the excess probability displays a
clear correlation as a function of galaxy type from ellipticals to irregulars,
which provide insights to the process of galaxy formation and the build up of
the Hubble sequence.  The resulting slope in the $\eta$-morphology relation
turns out to be positive in the sheet and void cases while it gets negative
for filaments and knots. These trends are consistent with the idea of a
continuous transition from irregular/spiral to elliptical morphology. The
direction of the transition is suggested by the gradual increase of the slope
as one follows the sequence knots--filaments--sheets--voids, i.e. from the
more to the less dense environments.  This indicates that spheroidal-like
systems are most probably formed in high-density regions as a result of
mergers between irregular/spiral objects. Conversely, the lower merger
probability in less dense environments permits the latter to retain their
morphology.

\section{Summary and Conclusions}
\label{sec:conc}

In this work we have presented a characterisation of the matter content of the 
LU including a cosmic web analysis and an environmental-galaxy morphology 
study based on a constrained cosmological simulation. This is only possible owing to 
the high-precision of the reconstructed initial conditions performed 
by the Bayesian self-consistent method of \cite{Kitaura13}. 
Our reconstruction covers a volume of $180^3\,h^{-3}\,{\rm Mpc^3}$ 
and it was produced using the spatial information provided 
by 31,107 2MRS galaxies \citep{Hess13}. To characterise 
the cosmic web of the reconstruction we have applied the tidal field tensor 
classification method on the nonlinear (T-web) and 
linear ({\rm div}\,V-web) density fields separately (see Section~\ref{sec:cosweb}). 
The latter has been computed using the reconstructed nonlinear velocity field 
within the linear theory approximation.

To estimate the expected cosmic variance level we used a suite 
of unconstrained $N$-body simulations with 
the same parameter setting as in our reconstructions. In this way, we 
were able to assess the consistency level of the specific LU realisation 
with the $\Lambda$CDM cosmology. To that end we have 
measured the corresponding VFFs and MFFs of voids, sheets, 
filaments and knots of the LU and compared them with those resulting 
from $N$-body universes with random initial seeds. 
This work brings up a number of novel aspects that are 
important to remark:

\begin{itemize}

\item[\checkmark] We use a state-of-the-art reconstruction of the nonlinear density 
and peculiar velocity fields based on high precision constrained simulations 
that reach an accuracy of about 3 Mpc. 
\\

\item[\checkmark] Based on such a reconstruction we present the first fully nonlinear 
cosmic web classification of the LU using the tidal field tensor method on 
the reconstructed density field.  
\\

\item[\checkmark] For the first time, we use the reconstructed nonlinear peculiar velocity 
field to characterise basic statistics of the LU cosmic web as well as 
to perform a study on the relation between galaxy morphology 
and environment. 
\\
\end{itemize}

In what follows we summarise the main findings of our work:

\begin{itemize}

\item We measured the DM density profile in the LU by using our 
$N$-body reconstruction. As a result, we found that the so-called `missing DM problem' 
can be simply interpreted as the consequence of ignoring matter located outside haloes above 
a given mass threshold. In fact, if we select simulated haloes with a similar virial mass 
cut as in the observations of \cite{Karachentsev12} (see also \citealt{Makarov11}) we are able 
to reproduce the observed value within $50\,\Mpc$ ($\Omega_{\rm M,LU}\approx0.1$) and the shape 
of the profile (see Fig.~\ref{fig:OmegaM_vs_r}). Instead, if we consider all matter present in the 
cosmic web, the matter density readily converges to the mean universal value. 
This suggests that the observationally-derived low matter density values cannot be the result of 
cosmic variance alone.  
\\

\item Both the reconstructed VFF and MFF statistics, computed using increasingly 
larger spheres centred on the observer's location, display, in general, 
a nice agreement with the expectations of $\Lambda$CDM. In particular, for 
a scale-radius larger than $r\gtrsim15\,\Mpch$ the deviations for all 
web environments are within $1\sigma$ of the expected cosmic variance. 
In particular, when considering a sphere with a radius of 
about $60\,\Mpch$, these statistics converge to the mean value of the random 
$\Lambda$CDM realisations, thus representing a {\it fair sample}. 
\\

\item We have measured the tendency of 2MRS galaxies to inhabit/avoid
  different environments as a function of morphology quantified in
  terms of the excess probability ratio $\eta$. In agreement with
  previous work, we found that elliptical systems are prone to be
  found in higher density regions, which we identify with knots and
  filaments in the web, rather than in sheets and voids.  The opposite
  is true for spirals. In general, we found that the $\eta$-morphology 
  correlations are best defined in real space where artificial shell crossing 
  effects caused by redshift-space distortions are not present.
\\ 

\item In particular, if the T-web classification is adopted to define the cosmic 
  web in our real-space reconstruction, elliptical galaxies show a clear signal
  ($\eta=1.27\pm0.05$) to preferentially reside in clusters, at a
  $5.3\sigma$ level, as opposed to sheets ($\eta=0.93\pm0.03$) and
  voids ($\eta=0.75\pm0.03$), at a level of $2.8\sigma$ and
  $9.8\sigma$, respectively.  Interestingly, we also found that
  elliptical galaxies show an excess probability of $\eta=1.11\pm0.03$
  in filaments that represents a $4.3\sigma$ detection. 
  However, filaments in the tidal field classification are not very
  well defined in the vicinity of high density regions, as it is shown in
  \citet[][]{Forero09} (their Fig.~1). Therefore, the latter result
  should be taken with caution. For spiral galaxies, we found a tendency to
  reside in voids ($\eta=1.17\pm0.03$) and sheets ($\eta=1.05\pm0.02$)
  at a $5.4\sigma$ and $2.1\sigma$, as opposed to filaments
  ($\eta=0.92\pm0.02$) and knots ($\eta=0.82\pm0.03$), at a
  $3.9\sigma$ and $5.6\sigma$ level, respectively. If, instead, we use
  the {\rm div}\,V-web classification to define the web we found that
  we can reproduce globally the same trends.
  \\

\item Irrespectively of the classification adopted to define the web we found that, 
in general, lenticular (S0) galaxies in the 2MRS catalogue do not show a preference 
for any particular environment. Irregulars, on the contrary, show similar trends 
as in the spiral case, although at much lower significance as a result of the small 
galaxy number in our sample. 
\\

\end{itemize}

As a final remark, we would like to note that our results concerning the validity of 
$\Lambda$CDM have to be taken as a consistency check since the 
constrained simulations were performed assuming that very same model. 
It is nevertheless remarkable that our constrained simulations are able to provide an explanation 
to the properties of the LU by either cosmic variance or observational biases 
without invoking a paradigm shift in cosmology. In this regard, we want to emphasise 
that the phases of the primordial 
fluctuations in our reconstructions -- which determine the location of peaks and troughs -- 
are purely constrained by the data. 
This work demonstrates that high precision constrained simulations 
can indeed help to characterise the properties of the LU in different 
environments. Therefore, we anticipate a large number of applications 
in which galaxy formation can be tested by directly cross-correlating 
observations to simulations including the full phase-space information.

\section*{Acknowledgments}
The authors acknowledge the anonymous referee for a constructive 
report that helped to improve this paper. SEN and FSK also thank Marius Cautun and Peter Creasey 
for useful comments on the manuscript. 
SEN, VM and SH acknowledge support by the Deutsche Forschungsgemeinschaft 
under the grants NU 332/2-1, MU1020 16-1 and GO563/21-1. NIL is also supported by the Deutsche 
Forschungsgemeinschaft. The simulations analysed in this work have been 
performed at NIC (J\"ulich, Germany). 

{
\small
\bibliographystyle{mn2e}
\bibliography{main}
}

\end{document}